\documentclass{jpp}

\usepackage{graphicx}
\usepackage{epstopdf}
\usepackage{xcolor}
\usepackage{amsmath}
\usepackage{hyperref}
\usepackage[export]{adjustbox}
\usepackage{float}

\usepackage{xspace}
\newcommand\ie{\textit{i.e.}\xspace}
\usepackage{comment}

\def \epsE {$\epsilon_{\mathrm{eff}}$}   

\usepackage{unicode}
\newcommand{\csso}{\fontencoding{LECO}\selectfont\char215}
\newcommand{\iotaslash}{\hspace*{0.1em}\iota\hspace*{-0.45em}\text{\csso}}

\title{Helicity of the magnetic axes of quasi-isodynamic stellarators}
\author{Katia Camacho Mata \aff{1}\corresp{\email{katia.camacho@ipp.mpg.de}}, Gabriel G. Plunk\aff{1} 
}
\affiliation{\aff{1}Max-Planck-Institut f\"ur Plasmaphysik, EURATOM Association, 17491 Greifswald, Germany
}

\begin{document}
\maketitle

 In this study, we explore the influence of the helicity of the magnetic axis —defined as the self-linking number of the curve— on the quality of quasi-isodynamic stellarator-symmetric configurations constructed using the near-axis expansion method \citep{camacho2022direct, plunk2019direct}. A class of magnetic axes previously unexplored within this formalism is identified when analyzing the axis shape of the QIPC  configuration \citep{subbotin2006integrated}: the case of half-helicity (per field period). We show these shapes are compatible with the near-axis formalism and how they can be used to construct near-axis stellarators with up-to 5 field-periods, \epsE{}$\approx 1.3\%$, and similar rotational transform as existing conventionally optimized designs, without the need of a plasma boundary optimization.

\section{Introduction}
 
 Quasi-isodynamic (QI) stellarators are relevant fusion reactor candidates due to their inherent steady-state operation, and reduced toroidal current, which prevents current-driven instabilities and reduces the Shafranov shift \citep{Helander_2009,Helander_2011,helander2014theory}. Additionally, if the so-called maximum-J condition is imposed \citep{rosenbluth1968}, additional benefits are expected with respect to turbulent transport, including trapped particle mode stability  \citep{proll2012resilience,helander2013collisionless,Helander_2015,Alcuson_2020} and some degree of ITG stabilization \citep{proll2022turbulence}. Stellarators have sometimes been considered to be at a disadvantage with respect to tokamaks due to the technical difficulties involved in their construction, but the success of the Wendelstein 7-X experiment has shown that these devices can be built and operated \citep{pedersen2018first, beidler2021demonstration}, placing the QI stellarator at the forefront of contenders for next-generation stellarator reactor concepts.

 Quasi-isodynamic configurations are a subset of omnigenous magnetic fields with poloidally closed contours of the field strength. Omnigenity requires the average radial drift of a trapped particle to vanish
 \begin{equation}
     \int (\mathbf{v_d} \cdot \mathbf{\nabla}\psi)dt = 0,
 \end{equation}
 where $\mathbf{v_d}$ is the guiding-centre drift velocity and the integration is performed over the bounce time of the trapped-particle orbit \citep{Hall,Cary,helander2014theory}. This guarantees radially confined collisionless orbits for trapped particles, which is in general not guaranteed in stellarators. Another subset of omnigenity is the case of quasi-symmetry, for which the intensity of the magnetic field is symmetric when expressed in magnetic coordinates. 
 
 Finding stellarator configurations consistent with omnigenity is an arduous task that is traditionally achieved through numerical optimization of the plasma boundary shape. The solution space of stellarators is vast, approximately an order of magnitude more degrees of freedom than tokamaks \citep{boozer2015stellarator}, and as a result, optimization is numerically expensive and highly dependent on the choice of good initial points. 
 
 The near-axis expansion (NAE) method was originally proposed by \cite{mercier1964equilibrium} and adapted for Boozer coordinates by \cite{garren1991magnetic}. This made it possible to construct stellarators that are omnigenous at least close to the axis without the need of an optimization procedure, as shown in \cite{landreman2018direct} and  \cite{landreman2019direct} for the case of quasi-symmetry and in \cite{plunk2019direct} and \cite{camacho2022direct} for quasi-isodynamic fields. We will describe the main equations of the NAE formalism in section \ref{NAE} and use it throughout this work to shed some light into the structure of the QI solution space. 

 An important aspect of the NAE is the intrinsic parametric nature of the construction. Instead of indicating the shape of the plasma boundary, the input parameters are more closely related to physically relevant parameters of the configuration, like the shape of the magnetic axis, directly contributing to the rotational transform $\iotaslash$, and the intensity of the magnetic field along the axis. By carefully choosing these parameters, a low level of neoclassical confinement can be achieved even at considerable distances from the axis, \ie{} at low aspect ratios \citep{camacho2022direct}. Optimization within these parameters has resulted in remarkably low neoclassical transport, as measured by the effective neoclassical ripple metric \epsE{} \citep{Jorge2022}. Boundary optimization using NAE configurations as initial points has also allowed the discovery of novel QI stellarators with excellent fast-particle confinement \citep{goodman2022constructing}.
 However, these methods have only led to configurations with a small number of field periods ($N{<}3$), in stark contrast to traditionally optimized QI configurations, like W7-X, QIPC \citep{subbotin2006integrated} or \cite{shafranov2004results} with $N{=}5,6,$ and 9, respectively.    
  
 The space of quasi-isodynamic configurations has proven to be complex and difficult to explore, as is evident from the tendency of traditional boundary optimizations to get stuck in local minima. The NAE has the potential to help overcome this issue, for example by mapping a suitably defined solution space, as been successfully done for the space of quasi-symmetric configurations in  \cite{landreman2022mapping} and  \citep{rodriguez2023constructing,rodriguez2022phases}. The magnetic axis shape, a three-dimensional curve, is one of the most important parameters for the NAE construction. The helicity of the axis, also called self-linking number, counts the number of times the normal Frenet vector rotates around the axis, and as discussed in \cite{rodriguez2022phases} and \cite{landreman2012omnigenity}, divides the space of symmetric configurations in quasi-axisymmetric and quasi-helical, with the transition between these phases corresponding to the QI case. In section \ref{Helicity} we expand on the definition of helicity, and the particularities of calculating it for the case of QI near-axis (NA) configurations. We also discuss the divisions in the solution space due to helicity and the impacts of this division in neoclassical confinement and rotational transform of the  configurations. 
 
 In section \ref{Half-helicity} we introduce the possibility of having fractional values of helicity and the conditions required on the curvature of the axis to attain such values. We proceed by showing that stellarator-symmetric configurations consistent with the NAE for QI can be constructed around axes with half-helicity, and we present a 2-field-period example with similar properties as the configuration described in \cite{camacho2022direct}, showing half-helicity configurations can be as good as their integer-helicity counterparts.   

We explore the implications of half-helicity (or more precisely, helicity of $n{+}\frac{1}{2}$ with $n$ any integer) in the QI space and show that this fractional helicity is the only one possible for the case of stellarator-symmetric NA configurations and corresponds to the transition between zero and integer helicities, making it a smaller solution space than the integer-helicity space. We show in section \ref{5FieldPeriods} that searching in this smaller space allows us to find configurations with higher number of field-periods and low \epsE{}. A 5-field-period example is shown as well as the details on the construction parameters' choice are discussed, including how to reduce elongation of the plasma boundary cross-section.

 \section{Near-axis expansion}\label{NAE}

The properties of a magnetohydrodynamic (MHD) equilibrium can be described in the vicinity of the magnetic axis by performing a Taylor expansion in the parameter $\epsilon = a/R$, the inverse aspect ratio, where $a$ and $R$ are the minor and major radius respectively. This procedure, using Boozer coordinates ($\psi,\theta,\varphi$), was shown in \citep{garren1991magnetic} to allow for the numerical construction of quasi-symmetric configurations, and was extended to consider quasi-isodynamic configurations at first order in \citep{plunk2019direct}. We will restrict our study to the case of first order, quasi-isodynamic, stellarator symmetric equilibria as described in \citep{camacho2022direct}. 

This method allows us to directly construct QI equilibria at first order by prescribing the shape of a magnetic axis $x_0$, the intensity of the magnetic field on the axis $B_0$, the number of field periods N, two toroidal geometric functions $\alpha(\varphi)$ and $d(\varphi)$, which will be later discussed in more detail, and a distance from the axis at which we desire to construct the equilibrium. In this formalism, the magnetic field intensity is given by
\begin{equation}\label{eq:magneticField}
    B(\epsilon, \theta, \varphi) \approx B_{0}(\varphi) + \epsilon B_{1}(\theta,\varphi)= B_{0}(\varphi) \left(1 + \epsilon d(\varphi) \cos[\theta - \alpha(\varphi) ] \right),
\end{equation}
and the plasma boundary is described by
 \begin{equation}\label{eq:NA_Boundary}
    \mathbf{x} \approx \mathbf{r}_{0} + \epsilon \left( X_{1}\mathbf{n}^{s} + Y_{1}\mathbf{b}^{s} \right),
\end{equation}
 with 
 \begin{gather}
    X_{1} = \frac{d(\varphi)}{\kappa^{s}} \cos{[\theta - \alpha (\varphi) ]} \\
    Y_{1} = \frac{2\kappa^{s}}{B_{0}(\varphi)d(\varphi)} \left( \sin{[\theta - \alpha (\varphi)]} + \sigma(\varphi) \cos{[\theta - \alpha (\varphi) ]}   \right). \label{eq:Y1}
\end{gather}
$\kappa$ and $\tau$ are the curvature and torsion of the magnetic axis and the quantity $\sigma(\varphi)$ can be found by solving 
\begin{equation}\label{eq:sigma}
    \sigma^{'} + (\iotaslash -\alpha^{'})\left(\sigma^{2}+ 1 +\frac{B_{0}^{2}{\Bar{d}}^{4}}{4} \right)-G_{0}\Bar{d}^{2}\left(\tau + I_{2}/2\right) = 0,
\end{equation}
self-consistently with the rotational transform on axis  $\iotaslash$. Here, $\Bar{d}(\varphi) = d(\varphi) /\kappa^{s}(\varphi)$, primes denote derivatives with respect to the toroidal Boozer angle $\varphi$. $G_{0}$ and $I_2$ are related to the first non-zero terms of the poloidal and toroidal current functions on axis, both can be assumed to be zero for QI configurations. 

The function $d(\varphi)$ needs to be specified as an input for the construction, but the choice is not entirely free; a set of conditions need to be fulfilled to be consistent with omnigenity and stellarator symmetry. It must vanish at all extrema of the on-axis magnetic field $B_0$
\begin{equation}
    d(\varphi_{\mathrm{min}})=d(\varphi_{\mathrm{max}})=0. 
\end{equation}
Given how the definition of the boundary depends directly on the quantity $\bar{d}$, the curvature must necessarily have zeros at the same toroidal locations and of the same order as $d$, to ensure a smooth plasma boundary. Additionally $d$ must be an odd function about $\varphi_{\mathrm{min}}$, which implies this function will always have derivatives of odd order equal zero at such points. Then $\kappa$ must also be zero up to odd order. The simplest case corresponds to both functions having first order zeros at the minima of $B_0$.

Finally, the last ingredient for constructing a NA QI equilibrium is the function $\alpha(\varphi)$. Just like $d$, it must be specified in a way consistent with omnigenity, albeit as shown in \citep{plunk2019direct} this condition is not compatible with a periodic plasma boundary. For this reason, omnigenity must be broken around the maxima of $B_0$. The method we will use is that described in section 4 of \citep{camacho2022direct}
\begin{equation}\label{eq:alpha_final}
    \alpha(\varphi) = \iotaslash (\varphi-\varphi_{\mathrm{min}}^{i})+\pi (2 m i + \tfrac{1}{2}) + \pi \left( m - \iotaslash /N \right) \left(\frac{\varphi-\varphi_{\mathrm{min} }^{i}}{\pi/N} \right)^{2k+1},
\end{equation}
where $m$ is the axis helicity, which the next section is devoted to, and $k$ is an integer that controls the size of the region in which the omnigenity conditions are fulfilled; increasing  $k$ results in a larger toroidal domain in which $\alpha$ is omnigenous but at the cost of a sharp behaviour around the maxima to maintain a periodic behaviour. 

\section{Helicity}\label{Helicity}
 
 The geometric properties of a 3D space curve $\mathbf{r}(\ell)$ can be described by the Frenet-Serret vectors, $(\mathbf{t},\mathbf{n},\mathbf{b})$. These orthogonal unitary vectors, named tangent, normal and binormal, respectively, are defined by
 \begin{equation*}
    \mathbf{t}(\ell) 
    = \frac{\mathbf{r}^'(\ell)}{||\mathbf{r}^'(\ell)||}, 
 \end{equation*}

 \begin{equation}\label{eq:FS_vectors}
    \mathbf{n}(\ell) = \frac{\mathbf{t}^'(\ell)}{||\mathbf{t}^'(\ell)||}, 
 \end{equation}
 
 \begin{equation*}
    \mathbf{b}(\ell) = \mathbf{t}(\ell) \times \mathbf{n}(\ell), 
 \end{equation*}
and are related to the curvature and torsion ($\kappa, \tau$) of the curve through the Frenet-Serret formulas: 
  \begin{eqnarray}\label{eq:Frenet-Serret formulas}
   \frac{d\mathbf{t}}{d\ell} \;=&    ||\mathbf{r}^'(\ell)|| \kappa \mathbf{n}  , \nonumber \\
   \frac{d\mathbf{n}}{d\ell}\; =& - ||\mathbf{r}^'(\ell)||\left[ \kappa\mathbf{t} + \tau\mathbf{b} \right], \\
    \frac{d\mathbf{b}}{d\ell} \;=& - ||\mathbf{r}^'(\ell)||\tau \mathbf{n} \nonumber.
\end{eqnarray}
 
 This frame plays an important role in the construction of exactly omnigenous stellarators at first order around the magnetic axis as described in \citep{landreman2018direct} and \citep{plunk2019direct}, as they are used to describe the plasma boundary as seen in equation \ref{eq:NA_Boundary}. 
 
 One interesting property of a magnetic axis described using this apparatus is the so-called helicity, $M$, the number of times the normal vector $\mathbf{n(\ell)}$ rotates poloidally around the axis after one toroidal turn. In this work we will often refer to the per-field-period helicity $m=M/N$, for example, in equation \ref{eq:alpha_final}, the definition of the $\alpha$ function. In the case of quasi-symmetric configurations, the helicity effectively divides the space of solutions as shown in \citep{landreman2018direct} and \citep{rodriguez2022phases}: a value of $M{=}0$ corresponds to axisymmetric configurations while an integer value of $M{\neq}0$ corresponds to quasi-helically symmetric solutions. 
 
\begin{figure}
\centering
  \includegraphics[width=0.74\textwidth]{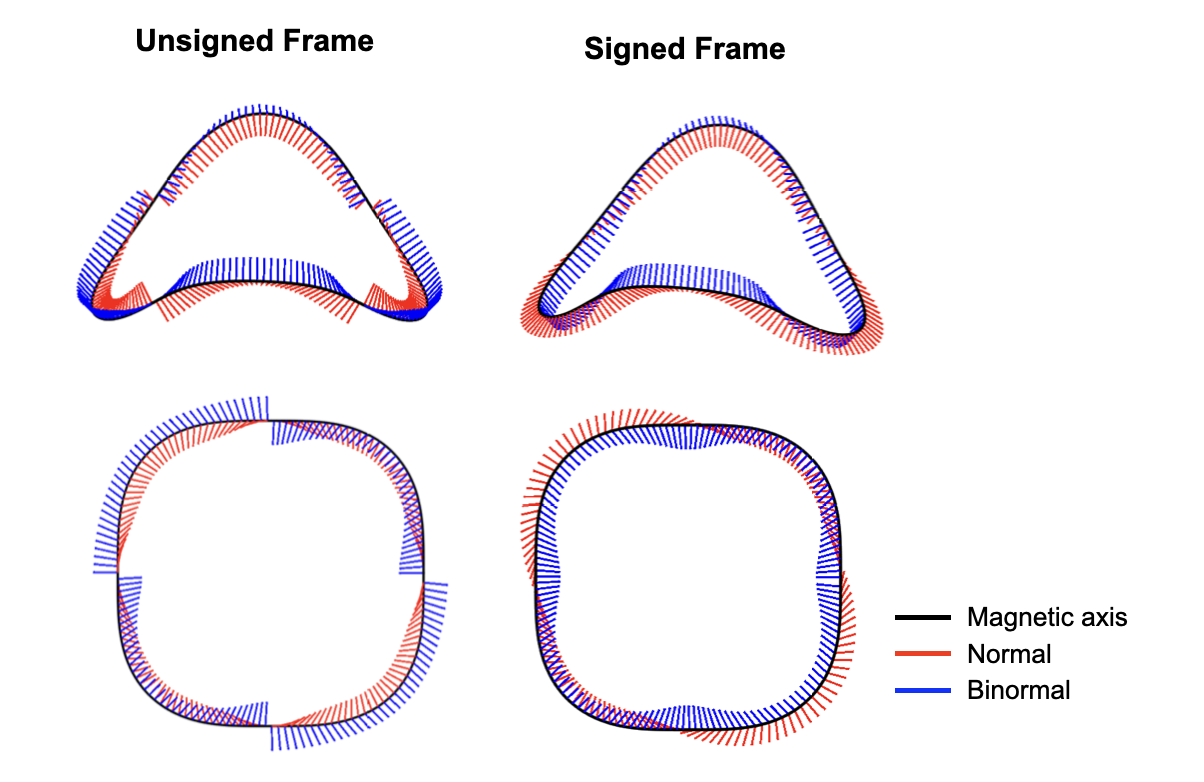}
\caption{(left) Unsigned Frenet frame, and (right) signed Frenet frame for a 2 field-period axis.Side view and top view are shown. Normal $\hat{n}$ and binormal $\hat{b}$ vectors are shown in red and blue, respectively. We can see the discontinuities in the frame are alleviated when using the signed curvature, $\kappa^s$.}
\label{fig:signedFrame}
\end{figure}

 For quasi-isodynamic configurations, which are the main focus of this work, calculating helicity requires extra considerations. As shown in the previous section, points of zero curvature are required in magnetic axes consistent with quasi-isodynamicity in the near-axis formalism. But we can see from the first equation in \ref{eq:Frenet-Serret formulas} that $\kappa = 0$ results in an undefined normal vector. In the presence of points of first-order zero curvature\footnote{We refer as first order zeros to the points where a function $F$  has values $F(\phi) = 0$, and higher order zeros to points where $\frac{d^{n-1}F}{d\phi^{n-1}}=0$ up to n-th order}, the Frenet frame undergoes a jump-like rotation which makes the frame, specifically the normal and binormal vectors non-continuous \citep{aicardi2000self}, as can be seen in the left side of figure \ref{fig:signedFrame}. This discontinuity can be alleviated by using the signed curvature $\kappa^{s}= s \kappa$, where $s = \pm 1$ changes at each point of odd-order zeros of curvature. In this new frame \citep{carroll2013improving}, the sign of the traditional Frenet normal and binormal vectors $\mathbf{n}(\ell),\mathbf{t}(\ell)$ also changes when $\kappa = 0$ resulting in a continuous frame as shown on the right side of figure \ref{fig:signedFrame}. Helicity for QI configurations is hence calculated using this modified Frenet frame to avoid any discontinuities in the plasma boundary shape.

\begin{figure}
    \centering
    \includegraphics[width=0.6\textwidth]{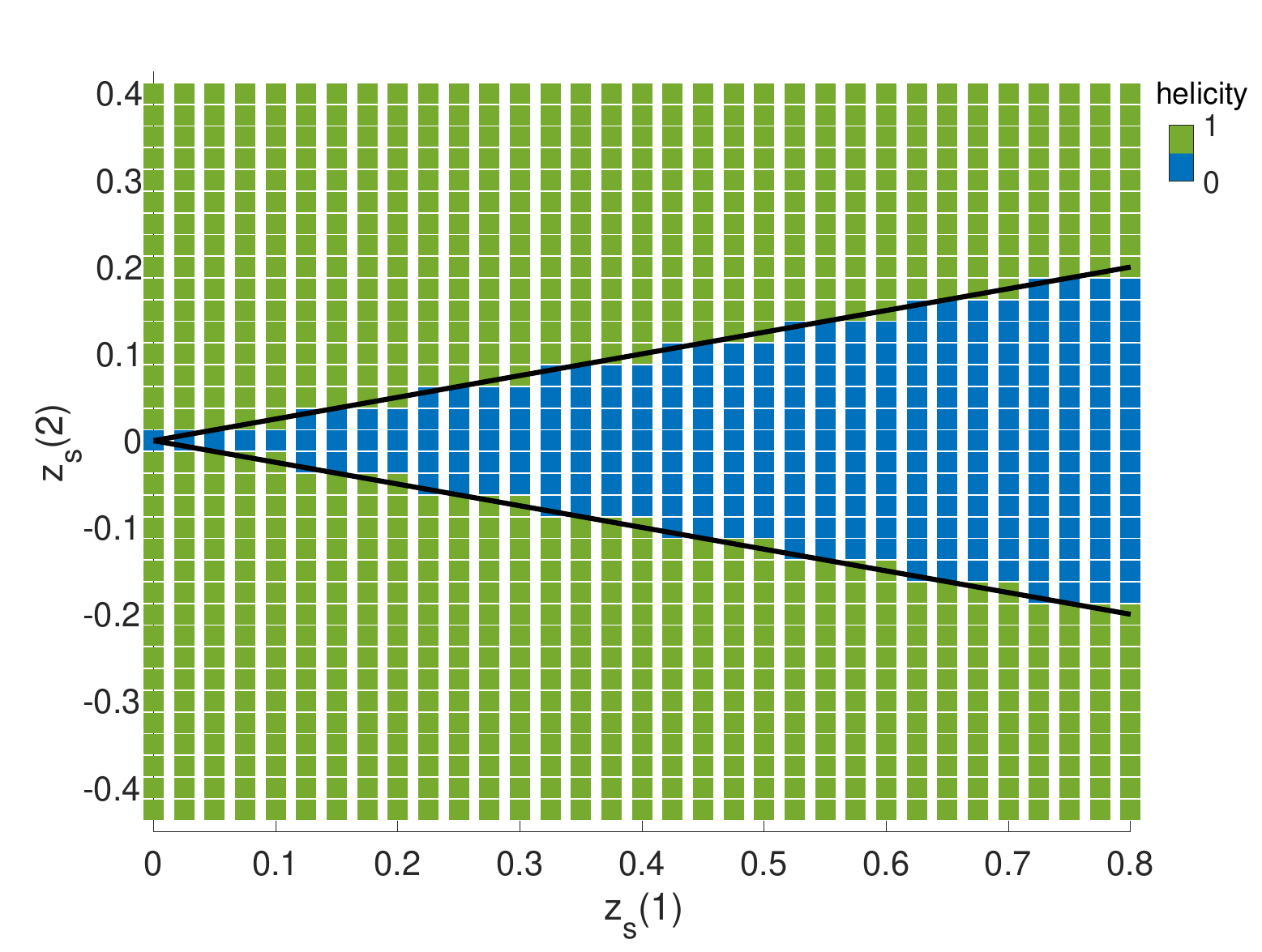}
    \caption{Space of 1 field-period near axis QI configurations and their helicity values. Each square corresponds to a NA QI configuration and its helicity value, green for $m{=}1$ and blue $m{=}0$. The solid lines show the transition between helicity regions and correspond to second order zeros of curvature at $\varphi_{\mathrm{min}}$ (upper line), and $\varphi_{\mathrm{max}}$ (lower line)} 
    \label{fig:helicity_1FP}
\end{figure}

 As in the case of quasi-symmetry, helicity also effectively divides the space of QI NA solutions as can be seen in figure \ref{fig:helicity_1FP}. Here, the helicity value for 1,156 one field-period, near-axis configurations is shown, each for a different axis shape while keeping the rest of the near-axis parameters constant. All these configurations have the same axis radial component
 \begin{equation}
     R(\phi) = 1 - 0.2 \cos (2\phi),
 \end{equation}
 which ensures two zeros of curvature are present at the minima and maxima of the magnetic field on axis. The vertical component of the axis shape is  different for each configurations and chosen as
 \begin{equation}\label{eq:Axis_2mode_Z}
     z(\phi) = z_s(1)\sin(\phi) + z_s(2)\sin(2\phi).
 \end{equation}
 Each square in figure \ref{fig:helicity_1FP} corresponds to a configuration with $z_s(1)$ ranging from 0 to 0.8 and $z_s(2)$ from -0.425 to 0.4, constructed using the near-axis method described in section \ref{NAE}. The aspect ratio is set to $A=20$, the function $\alpha(\varphi)$ is described by equation \ref{eq:alpha_final} with $k=2$, the magnetic field on axis is $B_0 = 1+0.15\cos (\varphi)$ and $d = 0.73 \kappa^s$. 
 
The solid black lines in figure \ref{fig:helicity_1FP}, bounding each helicity region, can be obtained analytically and correspond to the values of $z_s(1){-}z_s(2)$ that result in second order zeros of curvature,  \ie{} $\frac{d\kappa}{d\phi}|_{\varphi}=0$, at $\varphi_{\mathrm{min}}$ (upper line), and $\varphi_{\mathrm{max}}$ (lower line). The analytical derivation is shown in Appendix \ref{Appendix I}.
 
\begin{figure}
    \centering
    \includegraphics[width=0.9\textwidth]{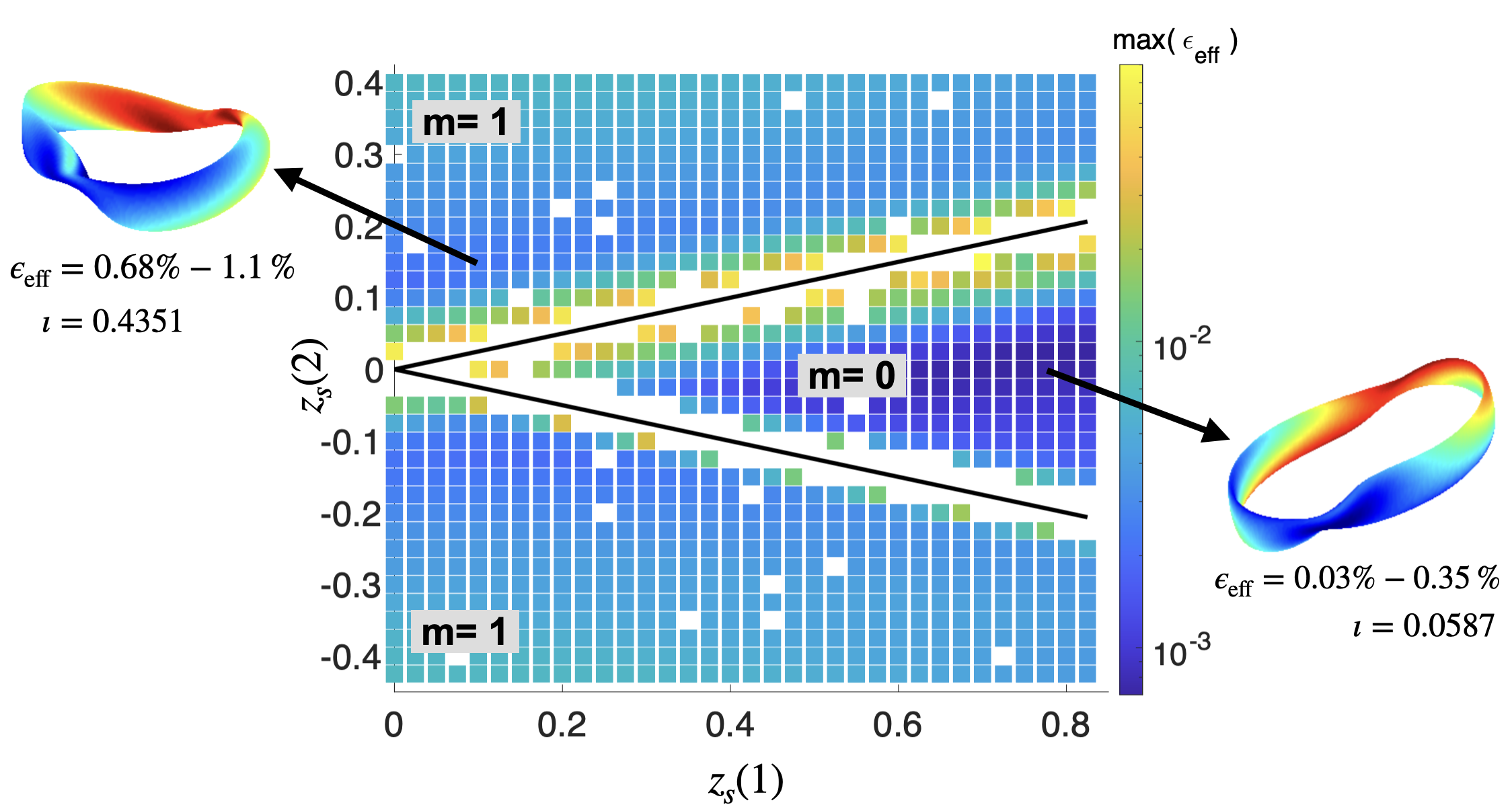}
    \caption{Space of 1 field-period, $A{=}20$, near axis QI configurations. $z_s(1)$ and $z_s(2)$ correspond to the parameters of the cylindrical Fourier representation of the axis Eq. \ref{eq:Axis_2mode_Z}. Each square is a near-axis configuration constructed with the parameters described in section \ref{Helicity}. Color coded is the maximum value of the effective ripple \epsE{}. Black solid lines correspond to the transition between regions with helicity 0 and 1. One configuration of each region is shown as an example of the properties in that space, the last closed surface and the intensity of the magnetic field on the boundary are shown. Zero-helicity configurations have lower \epsE{} but also lower rotational transform than their $m{=}1$ counterparts.} 
    \label{fig:helicity_1FP_eps_eff}
\end{figure}

 It is clear from figure \ref{fig:helicity_1FP} that the space of quasi-isodynamic solutions is divided into subregions by the helicity of the magnetic axis. But perhaps more surprising is the fact that this division is also present when calculating the effective ripple of each configuration in this space, as shown in figure \ref{fig:helicity_1FP_eps_eff}. Here, each of the  plasma boundary shapes of the configurations in figure \ref{fig:helicity_1FP} is used to solve for an MHD equilibrium using the VMEC code \citep{hirshman1983steepest}, and  $\epsilon_{\mathrm{eff}}$, the effective ripple, is calculated as described in \citep{drevlak2003effective}. The maximum value of this quantity is shown colour-coded in the figure. The white squares correspond to configurations for which VMEC could not find an equilibrium consistent with the provided plasma boundary. Once again, the division between regions of helicity is observed, and consistently lower values of $\epsilon_{\mathrm{eff}}$ and rotational transform $\iota$ are found for zero helicity configurations. One configuration from each region is also shown, but constructed at a smaller aspect ratio $A=10$, to exemplify these differences. On the left side, one corresponding to an $m=1$ axis, with a maximum $\epsilon_{\mathrm{eff}}=1.1\%$ and rotational transform $\iota = 0.4351$; while the configuration with $m=0$ has a substantially lower $\epsilon_{\mathrm{eff}}=0.35\%$ and  $\iota = 0.0587$.    
  
 Configurations with $m=1$ and remarkable low values of $\epsilon_{\mathrm{eff}}$ can be found as shown in \citep{Jorge2022} for one field-period, but it requires a significant effort in choosing the initial parameters. The lower values of effective ripple observed in the zero-helicity suggest that less effort might be necessary in this region of the parameter space. 
 
 Helicity seems to play an important role in the near-axis quasi-isodynamic space.  Although the integer helicity space seems to correlate low rotational transform and low $\epsilon_{\mathrm{eff}}$, this occurs because helicity is directly related to the integrated torsion of the magnetic axis, which is an important mechanism for rotational transform generation \citep{helander2014theory}. But optimized examples of QI configurations with low $\epsilon_{\mathrm{eff}}$ and higher per-field-period rotational transform exist. The exploration of such examples may shed light light on novel ways of generating near-axis equilibria with higher $N$ and small $\epsilon_{\rm eff}$. 
 

\section{Half helicity}\label{Half-helicity}

In the previous section, we restricted ourselves to the case of integer helicity, implying that the only case with less than complete rotation of the normal is the zero helicity case.  
 But it is also possible to construct curves with fractional values of per-field period helicity. 
 In fact, such behavior is exhibited by the magnetic axis of QIPC  \citep{subbotin2006integrated}, a quasi-isodynamic stellarator found through numerical optimization. This configuration has poloidally closed contours of the magnetic field close to the axis and an effective ripple profile that increases radially, making it more similar to a near-axis QI configuration than W7-X. 
 
\begin{figure}
\centering
  \includegraphics[width=0.9\textwidth]{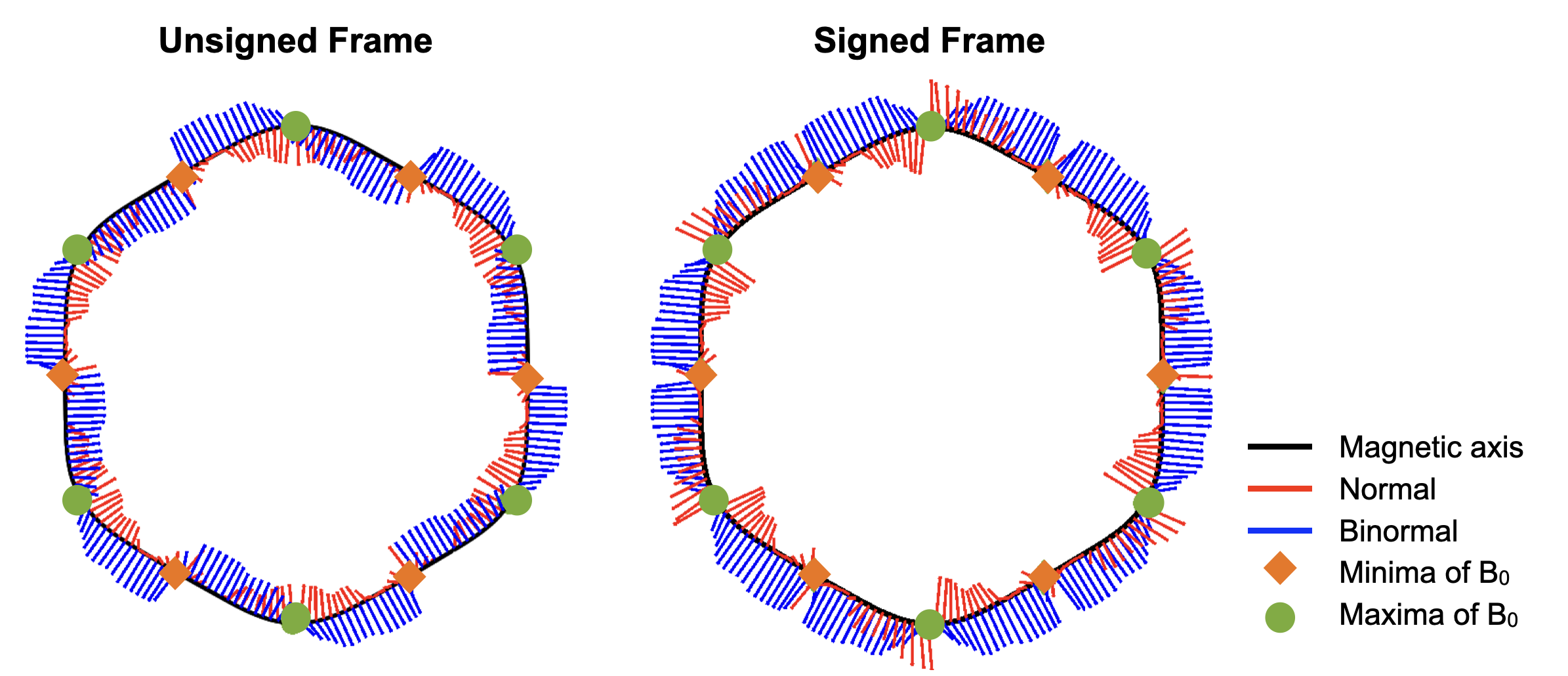}
\caption{(left) Unsigned and (right) signed Frenet frame for the magnetic axis of QIPC, a 6 field-period QI optimized configuration \citep{subbotin2006integrated}. Green circles indicate the toroidal locations of the maxima of $B_0$, marking to the beginning/end of a  magnetic field period. Green diamonds indicate the minima of $B_0$. Normal and binormal Frenet vectors are shown in red and blue lines, respectively. A discontinuity in the frame at minima of $B_0$ is evident in the signed frame from the direction change of the normal vector. This discontinuity is alleviated when using the signed frame and is transfered to the maxima of $B_0$. }
\label{fig:QIPC_Frame}
\end{figure}

 Although QIPC's magnetic axis does not have points of exactly zero curvature at the extrema of $B_0$, the curvature does reach relatively small values, and its signed Frenet frame behaves in a similar way to near-axis configurations, having a flip at points where the magnetic field on axis has its minimum. Figure \ref{fig:QIPC_Frame} shows this axis together with the normal and binormal vectors associated with its signed Frenet frame, the change in sign is applied at the position of the minima of $B_0$.  As we can see from the figure, the normal vector does not complete a whole rotation around the axis each field period, instead the normal vector seems to complete exactly a half rotation, ending at the opposite  direction it started at the beginning of the period. We note that all QI designs found in the past by numerical optimization share this half-rotation feature ($m=1/2$). However, this case has not been considered when constructing NA QI configurations in the literature. As we will show, it allows us to access a region in the solution space where good confinement can be achieved with a relatively high number of field periods.
 
 Let us now discuss the particularities of constructing a magnetic axis with $m=1/2$. We know a magnetic axis consistent with omnigenity needs to have points of zero curvature at all minima and maxima of the magnetic field \citep{plunk2019direct}, which requires  calculating helicity in the signed Frenet frame, since the axis helicity is otherwise not well defined. Additionally, for a curve to have half helicity, we require the normal vector to perform half a rotation per field period. This situation is not consistent with first order zeros of curvature at both $\varphi_{\mathrm{min}}$ and $\varphi_{\mathrm{max}}$, at least not for the case of analytic stellarator symmetric axis curves, which can be seen by re-examining the expansion near zero curvature points, as follows.

As shown in section 3 of \cite{camacho2022direct}, omnigenity conditions imposed in the near-axis expansion for stellarator symmetric configurations, require odd order zeros of $\kappa$ at the minima of $B_0$. If the axis is described as a Fourier series on the cylindrical $R$ and $z$ coordinates,
 \begin{eqnarray}
    R(\phi) = \sum_{n=0}^{n_\mathrm{max}} R_c(n) \cos(n N \phi),\label{eq:R-Fourier}\\
    z(\phi) = \sum_{n=1}^{n_\mathrm{max}} z_s(n) \sin(n N \phi).\label{eq:z-Fourier}
\end{eqnarray}
Then having a point of zero curvature at first order requires $\frac{d^{2}x}{d\phi^2}\cdot \hat{R} = 0$ at such point (Appendix II in \cite{camacho2022direct}), which given the definition of $\mathbf{n}$ (eq. \ref{eq:FS_vectors}), is equivalent to the condition
\begin{equation}\label{eq:r_dot_n}
    \mathbf{n}\cdot \hat{R}=0,
\end{equation} 
at $\varphi = \varphi_{\mathrm{min}}$. The case $m=\frac{1}{2}$ corresponds to the normal vector performing a $\pi$-rotation in the $\hat{R}-\hat{z}$ plane per field period. Hence, after half a period, at $\varphi = \varphi_{\mathrm{max}}$, a $\frac{\pi}{2}$-rotation would require the normal vector to point in the radial direction
\begin{equation}
    \mathbf{n} = \hat{R}.
\end{equation}
As shown above and explained in greater detail in Appendix \ref{Appendix I}, odd order zeros of curvature are generally inconsistent with a normal vector having a component in the radial direction, but consistent with even-order zeros.  We can intuitively visualize each increase in the order of the zeros of curvature as introducing a flip on the Frenet frame, hence the need for different behaviour of $\kappa$ at the minima and maxima of $B_0$ to preserve a discontinuity of the frame that is not corrected by the sign change described in previous section. We can then construct half-helicity closed curves, just by requiring a first order zero of curvature at $\varphi = \varphi_{\mathrm{min}}$ and a second order zero at $\varphi = \varphi_{\mathrm{max}}$. We note that these arguments rely on continuity of the derivatives of the curve, which can safely be assumed for curves constructed by Fourier series, but need not be the case more generally, as with curves found by solving Frenet-Serret equations directly.  We also remark that, when assuming stellarator symmetry as is done in this work, the only helicities possible are either integer or half-helicity, since this is the only fractional helicity allowing for points of zero curvature and continuity of the plasma boundary. Different fractional helicities might be possible for certain non-stellarator symmetric configurations, but this issue needs to be explored further.

 
 
 Following the same procedure described in  Appendix \ref{Appendix I}, we perform a local expansion of a stellarator-symmetric magnetic axis described by equations \ref{eq:R-Fourier} and \ref{eq:z-Fourier}, and impose conditions on the curvature of such axis, which we can transform into conditions for the Fourier coefficients $R_{c}(n)$ and $z_{s}(n)$. The local condition for having first order zeros of curvature is
 \begin{equation}\label{eq:1stOrderKappa}
     R_{2} = R_{0},
 \end{equation}
where $R_n$ and $z_n$ denote the $n$-th derivatives with respect to $\phi$. For a second order zero of $\kappa$, we additionally need to fulfill
  \begin{equation}\label{eq:2ndOrderKappa}
     z_{3} = 2z_{1}.
 \end{equation}

\begin{figure}
\centering
  \includegraphics[width=0.75\textwidth]{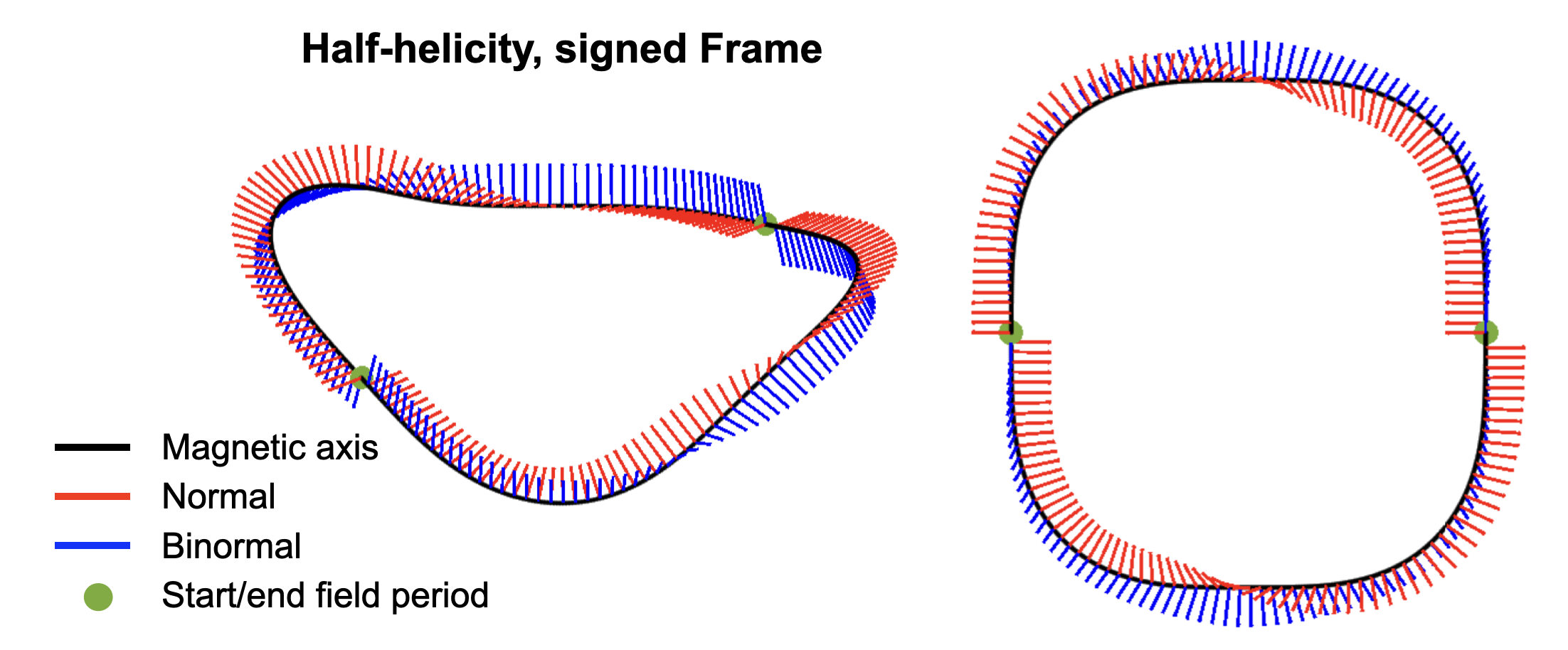}
\caption{ Signed Frenet frame for a half-helicity 2 field-period curve. Note the discontinuity that remains at the location where the magnetic field strength is taken to be maximal, where the normal vector points in the radial direction.}
\label{fig:HalfHelicity_Frame}
\end{figure}

We can apply the previous conditions to a truncated Fourier representation to obtain conditions on the Fourier coefficients that can be used to construct curves with the curvature properties required to obtain a half-helicity behaviour. We can consider, for example, the family of curves.
\begin{align}
    \label{eq:Axis_Truncated_FourierRepresentation}
     R =& 1 + R_c(1) \cos(N \phi) + R_c(2) \cos(2 N \phi) + R_c(3) \cos(3 N \phi), \nonumber\\
     z =& z_s(1) \sin(N \phi) + z_s(2) \sin(2 N \phi) + z_s(3) \sin(3 N \phi). 
\end{align}
Applying condition \ref{eq:1stOrderKappa} at $\phi=2\pi/N$ and both the previous condition and \ref{eq:2ndOrderKappa} to $\phi=0$ returns the following restrictions on the Fourier coefficients necessary to construct curves with half helicity per field period,
\begin{align}\label{eq:Axis_HalfHelicity_Conditions}
R_c(1) =& -\frac{R_c(3)+9N^2R_c(3)}{1+N^2},\\
R_c(2) =& -\frac{1}{4 N^2+1},\\
z_s(1) =& -\frac{4z_s(2)+8N^2z_s(2)+6z_s(3)+27N^2z_s(3) }{2+N^2}.
\end{align}
An axis constructed using this method is shown in figure \ref{fig:HalfHelicity_Frame}, where the parameters were chosen as $N=2$, $R_c(3) = 5 \times 10^{-3}$, $z_s(2) = 3 \times 10^{-2}$ and $z_s(3) = 1.5 \times 10^{-3}$. We can see the rotation of the signed normal and binormal vectors has the expected behaviour, with one jump per field period. 

\section{Constructing half-helicity near-axis stellarators}

Now we can proceed to use this axis to construct a  near-axis configuration, as described in Section \ref{NAE}. A natural question to ask is whether the discontinuity in the Frenet frame at maxima of the magnetic field, required for half-helicity magnetic axes, has a detrimental effect on the continuity of the plasma boundary. We confirm that the formalism developed in \citep{camacho2022direct} is valid for fractional helicity and results in continuous plasma boundaries. 

 We know from the previous section that the Frenet frame, specifically the normal and binormal vectors are discontinuous at maxima of $B_{0}$. Given how the plasma boundary of a near-axis expansion equilibrium is constructed using equation \ref{eq:NA_Boundary}
\begin{equation*}
    \mathbf{x} \approx \mathbf{r}_{0} + \epsilon \left( X_{1}\mathbf{n}^{s} + Y_{1}\mathbf{b}^{s} \right),
\end{equation*}
it would be natural to assume the discontinuity of $\mathbf{n}^{s}$ and $\mathbf{b}^{s}$ translates into a discontinuity of the boundary shape, but this is only true if $X_1$ and $Y_1$ are continuous. Let us analyze the behaviour of this first order corrections to the plasma boundary
\begin{gather}\label{eq:X0_Y0}
    X_{1} = \frac{d(\varphi)}{\kappa^{s}} \cos{[\theta - \alpha (\varphi) ]} \\
    Y_{1} = \frac{2\kappa^{s}}{B_{0}(\varphi)d(\varphi)} \left( \sin{[\theta - \alpha (\varphi)]} + \sigma(\varphi) \cos{[\theta - \alpha (\varphi) ]}   \right). 
\end{gather}

The quantity $\bar{d} = d(\varphi)/\kappa^{s}$ is, by definition, constructed to be symmetric and continuous at every point, in particular at $\varphi_{\mathrm{min}}$ and $\varphi_{\mathrm{max}}$ as shown in \citep{camacho2022direct}. The magnetic field on axis $B_0(\varphi)$ is also continuous at points where the Frenet frame is discontinuous. Hence any possible discontinuity in $X_1$ and $Y_1$ is going to depend on the behaviour of $\alpha(\varphi)$, described by
\begin{equation}
    \alpha(\varphi) = \iotaslash (\varphi-\varphi_{\mathrm{min}}^{i})+\pi (2 m i + \tfrac{1}{2}) + \pi \left( m - \iotaslash /N \right) \left(\frac{\varphi-\varphi_{\mathrm{min} }^{i}}{\pi/N} \right)^{2k+1}. 
\end{equation}

Note the first term ensures that the magnetic perturbation is approximately in phase at two bounce points (omnigenity), while the final term is designed to break this criterion away from minima of the field; see \cite{camacho2022direct} for more details. For half-helicity curves $m=1/2$ and let us consider the first field-period, then $i=1$ and $\varphi_{\mathrm{min}}^{i}=\pi/N$. Now let us see the value the $\alpha$-function takes at two consecutive maxima, \ie{} the beginning and end of a field-period, corresponding to $\varphi^{1}_{\mathrm{max}}=0$ and $\varphi^{2}_{\mathrm{max}}=2\pi/N$. 

\begin{equation*}
    \alpha(\varphi^{1}_{\mathrm{max}}) = \iotaslash (-\pi/N)+ \pi (1 + \tfrac{1}{2}) + \pi \left( \tfrac{1}{2} - \iotaslash /N \right) \left(\frac{-\pi/N}{\pi/N} \right)^{2k+1}= \pi,
\end{equation*}
\begin{equation*}
    \alpha(\varphi^{2}_{\mathrm{max}}) = \iotaslash (2\pi/N-\pi/N)+ \pi (1 + \tfrac{1}{2}) + \pi \left( \tfrac{1}{2} - \iotaslash /N \right) \left(\frac{\pi/N}{\pi/N} \right)^{2k+1}= 2\pi.
\end{equation*}

We can see that for $m=1/2$ the function increases by $\pi$ per field-period, contrary to the $2\pi$-periodicity of the integer helicity curves. Now let us see the impact this has on the functions $X_1$ and $Y_1$ describing the plasma boundary,

\begin{gather*}
    X_{1}(\theta,\varphi=\varphi^{1}_{\mathrm{max}}) = \frac{d(\varphi)}{\kappa^{s}} \cos{[\theta - \pi ]} = - \frac{d(\varphi)}{\kappa^{s}} \cos{[\theta]}, \\
    X_{1}(\theta,\varphi=\varphi^{2}_{\mathrm{max}}) = \frac{d(\varphi)}{\kappa^{s}} \cos{[\theta - 2\pi ]} =  \frac{d(\varphi)}{\kappa^{s}} \cos{[\theta]} = - X_{1}(\theta,\varphi=\varphi^{1}_{\mathrm{max}}), 
\end{gather*}

\begin{equation*}
\begin{split}
    Y_{1}(\theta,\varphi=\varphi^{1}_{\mathrm{max}}) &= \frac{2\kappa^{s}}{B_{0}(\varphi)d(\varphi)} \left( \sin{[\theta - \pi]} + \sigma(\varphi) \cos{[\theta - \pi]}   \right) \\ &= -\frac{2\kappa^{s}}{B_{0}(\varphi)d(\varphi)} \left( \sin{[\theta]} + \sigma(\varphi) \cos{[\theta]}   \right), \\ 
    Y_{1}(\theta,\varphi=\varphi^{2}_{\mathrm{max}}) &= \frac{2\kappa^{s}}{B_{0}(\varphi)d(\varphi)} \left( \sin{[\theta - 2\pi]} + \sigma(\varphi) \cos{[\theta - 2\pi]}   \right) \\ &= -Y_{1}(\theta,\varphi=\varphi^{1}_{\mathrm{max}}).
\end{split}
\end{equation*}
As shown above, the boundary components $X_1$ and $Y_1$ are discontinuous at maxima of $B_{0}$. This discontinuity is, necessarily, of the same type as that of the Frenet-frame; continuity of the coordinate mapping \ref{eq:NA_Boundary} is thus achieved despite a discontinuous signed Frenet frame. 
Now we can use the formalism developed in \citep{camacho2022direct} to construct a magnetic configuration that is quasi-isodynamic at first order, using a half-helicity magnetic axis. 

\subsection{A 2 field-period half-helicity NA stellarator} \label{sec:2FP_configuration}

Let us now proceed to show an example of a 2-field-period configuration constructed around a half-helicity magnetic axis

\begin{equation}\label{eq:R0_2FP}
    R = 1 - 0.074\cos{(2\phi)} - \tfrac{1}{17} \cos{(4\phi)} + 0.01 \cos{(6\phi)} ,
\end{equation}
\begin{equation}\label{eq:Z0_2FP}
    z =  -0.474 \sin{(2\phi)} + 0.06 \sin{(4\phi)} + 6\times10^{-3}\sin{(6\phi)}.
\end{equation}

\begin{figure}
\centering
   \includegraphics[width=0.49\textwidth]{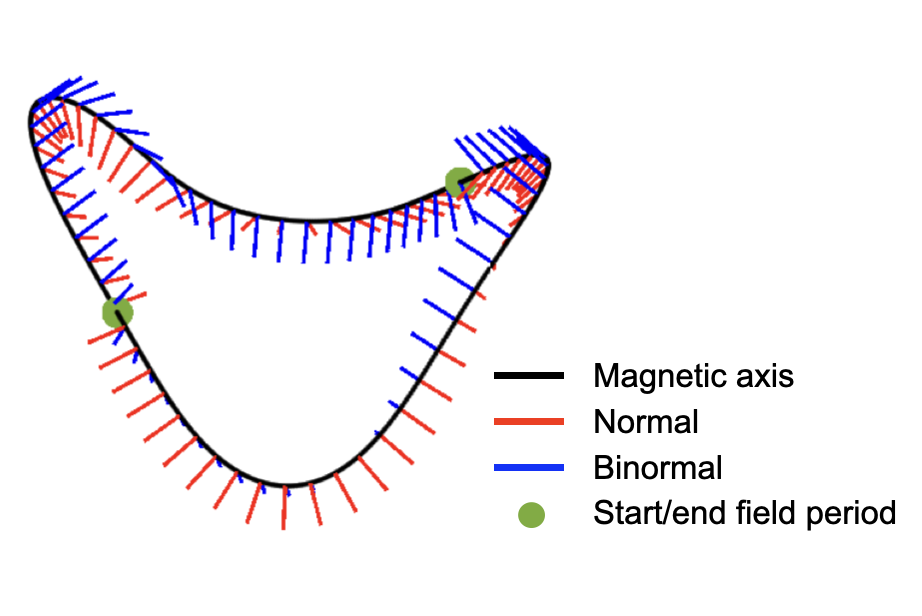} 
    \includegraphics[width=0.49\textwidth]{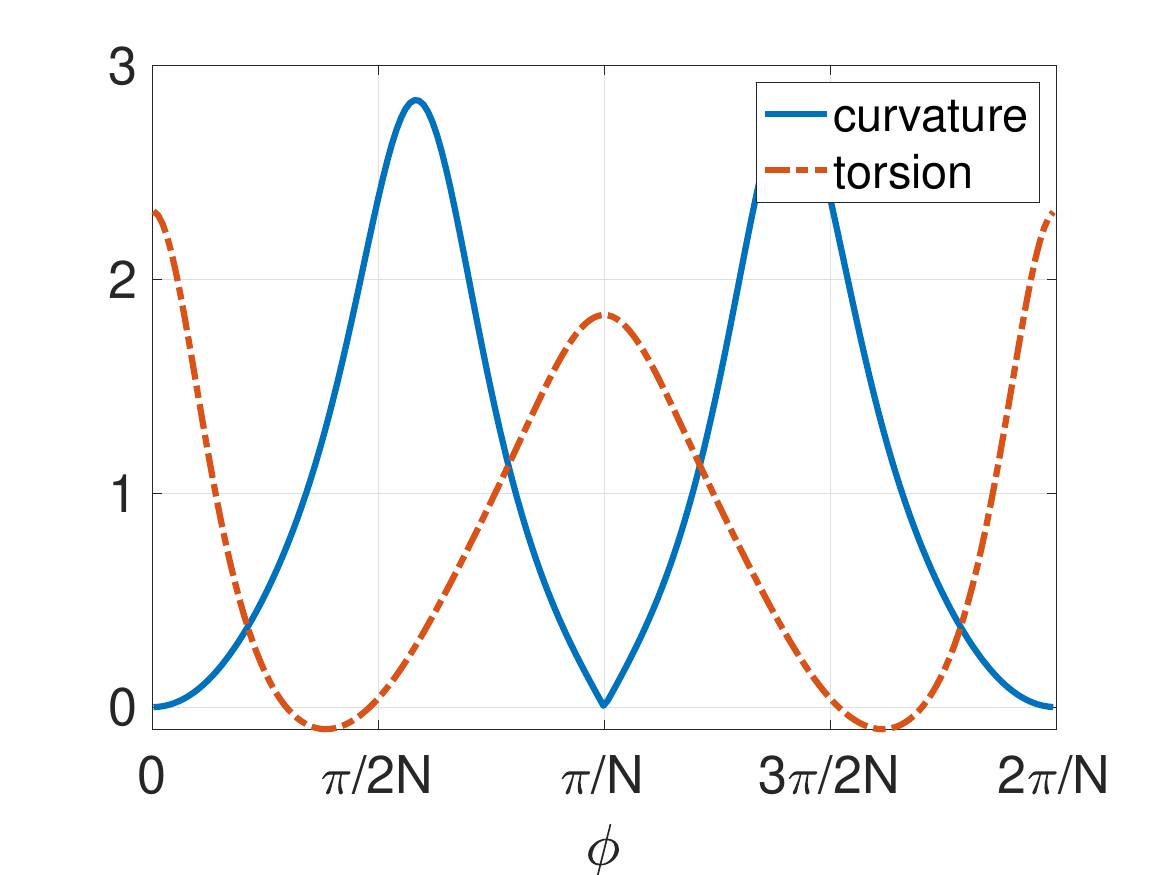}
\caption{(left) 2 field-period half-helicity magnetic axis described by expressions \ref{eq:R0_2FP} and \ref{eq:Z0_2FP}. The normal and binormal vectors are shown, the discontinuity of the Frenet frame is evident at the beginning/end of the period. (right) Curvature (solid) and torsion (dashed)  per-field-period profiles of the axis. Note the first-order zero of curvature at $\phi {=} \pi / N$ and second-order zeros at  $\phi{=} 0, 2\pi /N $, necessary to achieve values of helicity $m{=}\frac{1}{2}$.}
\label{fig:axis_2FP}
\end{figure}

This axis was constructed to have $m{=}1/2$ as described in section \ref{Half-helicity},  and the values of the NA parameters were chosen such that the corresponding VMEC equilibrium has small $\epsilon_{\mathrm{eff}}$ on its boundary, as discussed below. The shape of this magnetic axis, together with its signed Frenet frame, and curvature and torsion profiles are shown in figure \ref{fig:axis_2FP}, we can observe the expected curvature behaviour for half-helicity: a first-order zero at half-period and second-order zeros at the end/beginning of the period.    
 
The intensity of the magnetic field on this axis must also be specified and was chosen as
\begin{equation*}
    B_{0} (\varphi) = 1 + 0.165\cos{(2\varphi)},
\end{equation*}
which, as required by the near-axis formalism, has extrema at points where the curvature of the magnetic axis is zero. Another parameter playing an important role on the quality of the solutions, specifically in their elongation and maximum values of $\epsilon_{\mathrm{eff}}$ is the function $d(\varphi)$. It was chosen to be proportional to the curvature of the axis as $d=d_{\kappa}\kappa^s$, with
 \begin{equation*}
     d_{\kappa}(\varphi) = \sqrt{0.4/B_{0}(\varphi)}.
 \end{equation*}
 This particular form of $d(\varphi)$, inversely proportional to $B_0$, is useful for controlling elongation, and here helps flatten
 the elongation profile, \ie{} reduce its variation in $\varphi$, resulting in smaller maximum elongations than using a constant $\bar{d}$ choice. \footnote{Note that in this particular case $d_{\kappa}{=}\bar{d}$, but in the following examples, when more complicated $d(\varphi)$ expressions are used, this will not be generally true}   It is important to remember the elongation dependence on $\bar{d}$ is complicated and includes other quantities of the near-axis construction, making it difficult to predict which combination of parameters will result in a desired elongation behaviour. The freedom in $\bar{d}$, however, actually allows more control of the elongation profile, as compared to quasi-symmetric configurations  where it must be constant; this will be demonstrated in section \ref{5FieldPeriods}.   

\begin{figure}
\centering
  \includegraphics[width=0.99\textwidth]{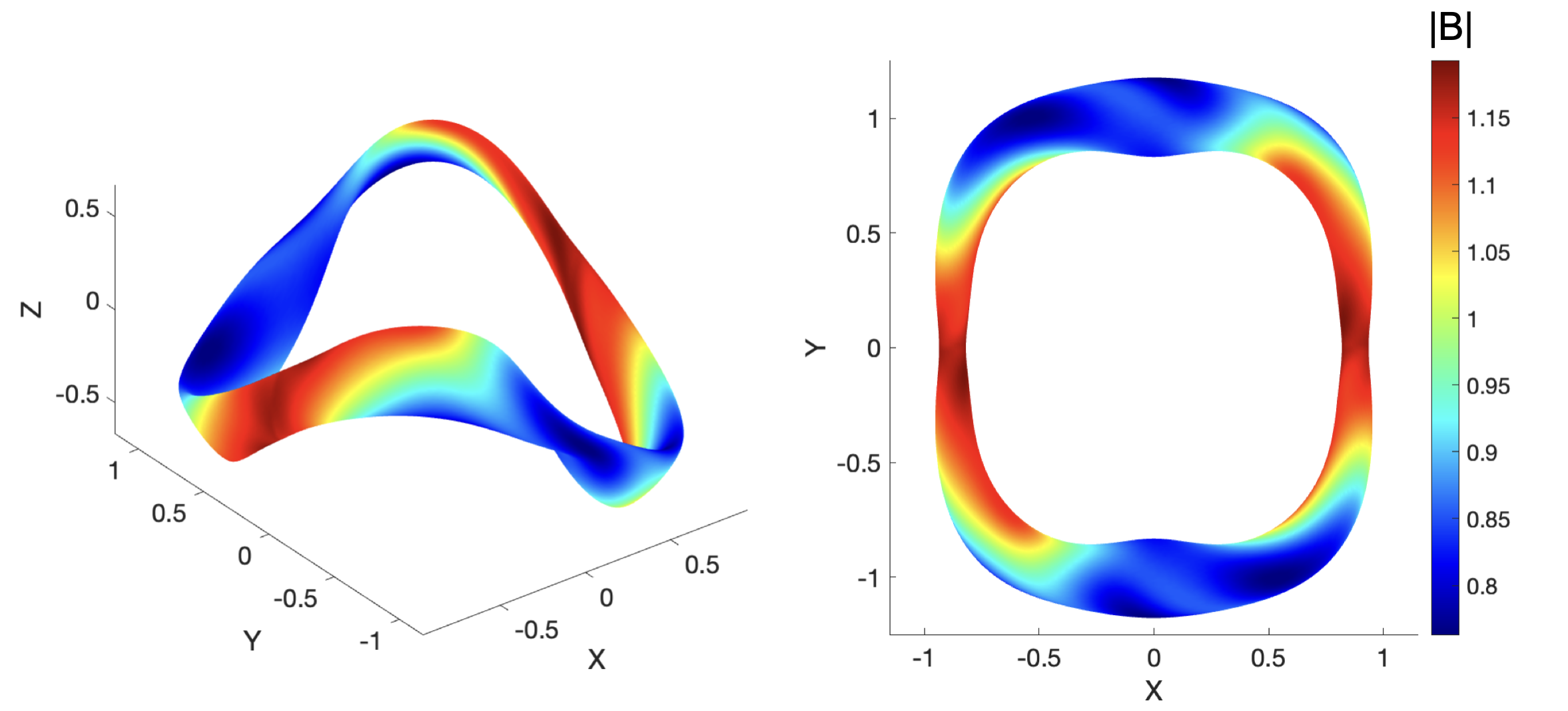}
\caption{ Magnetic field intensity in the plasma boundary for a 2 field-period half-helicity configuration. (left) Side view and (right) top view.}
\label{fig:2FP_Boundary}
\end{figure}

\begin{figure}
\centering
  \includegraphics[width=0.49\textwidth]{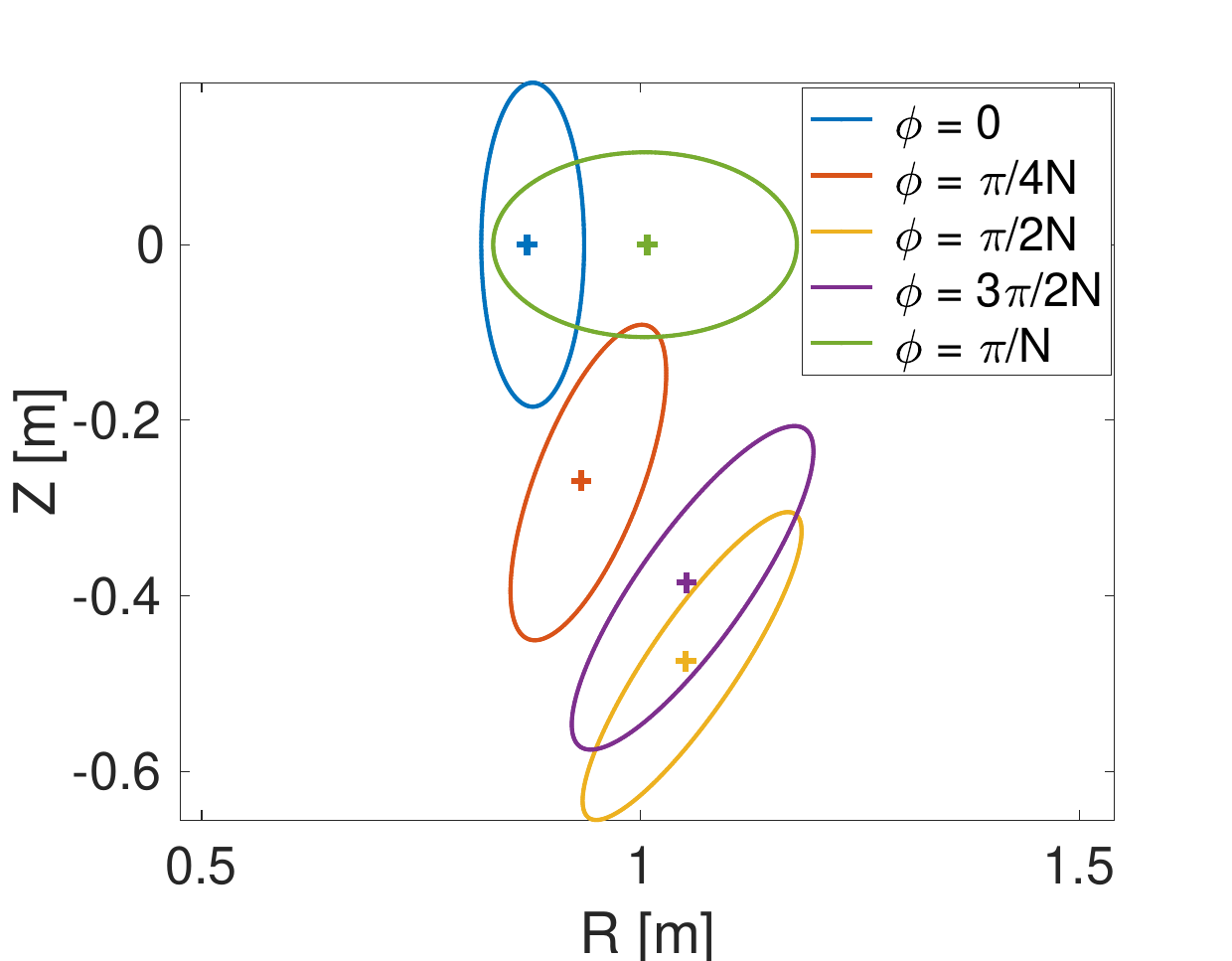}
   \includegraphics[width=0.49\textwidth]{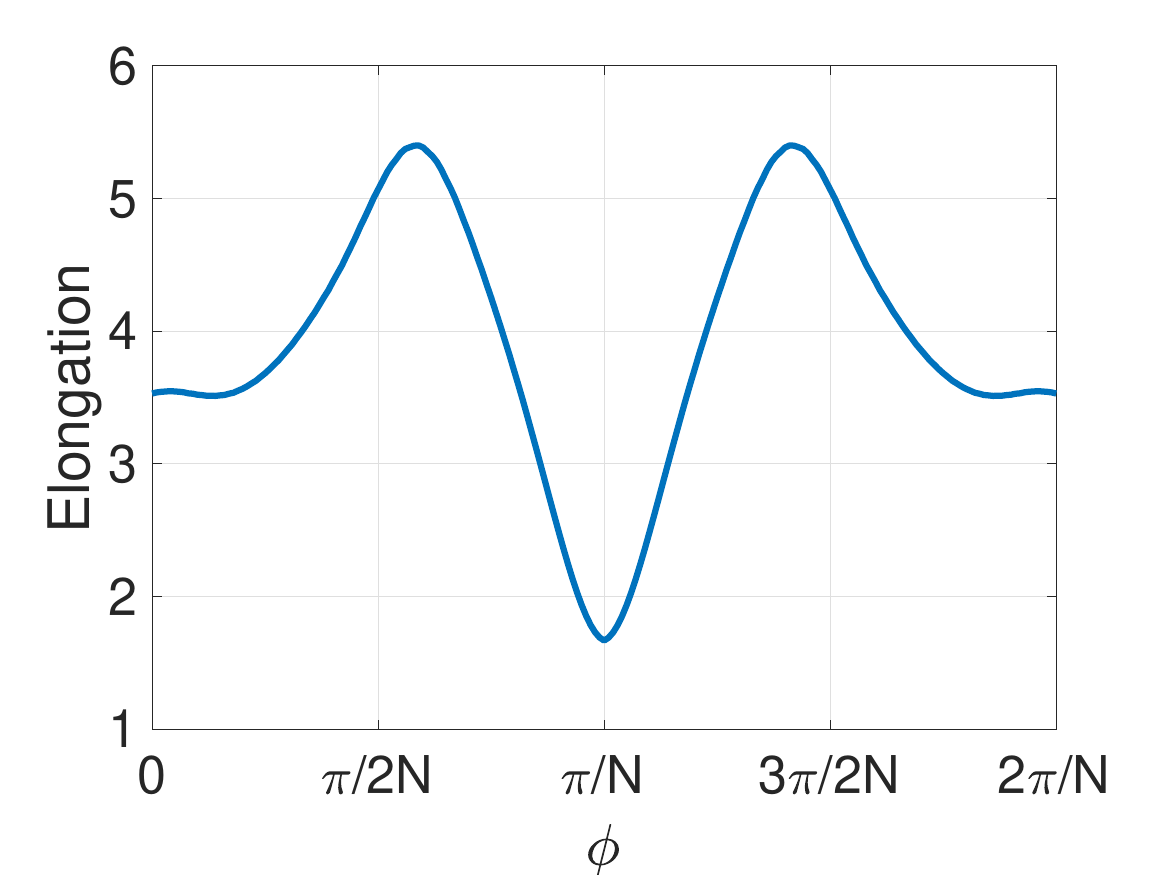} 
\caption{(left) Cross-sections of the 2 field-period configuration shown in figure \ref{fig:2FP_Boundary} at different toroidal values. (right) Elongation toridal profile for the same configuration.}
\label{fig:2FP_cross_sections_elongation}
\end{figure}

\begin{figure}
\centering
\includegraphics[width=0.32\textwidth]{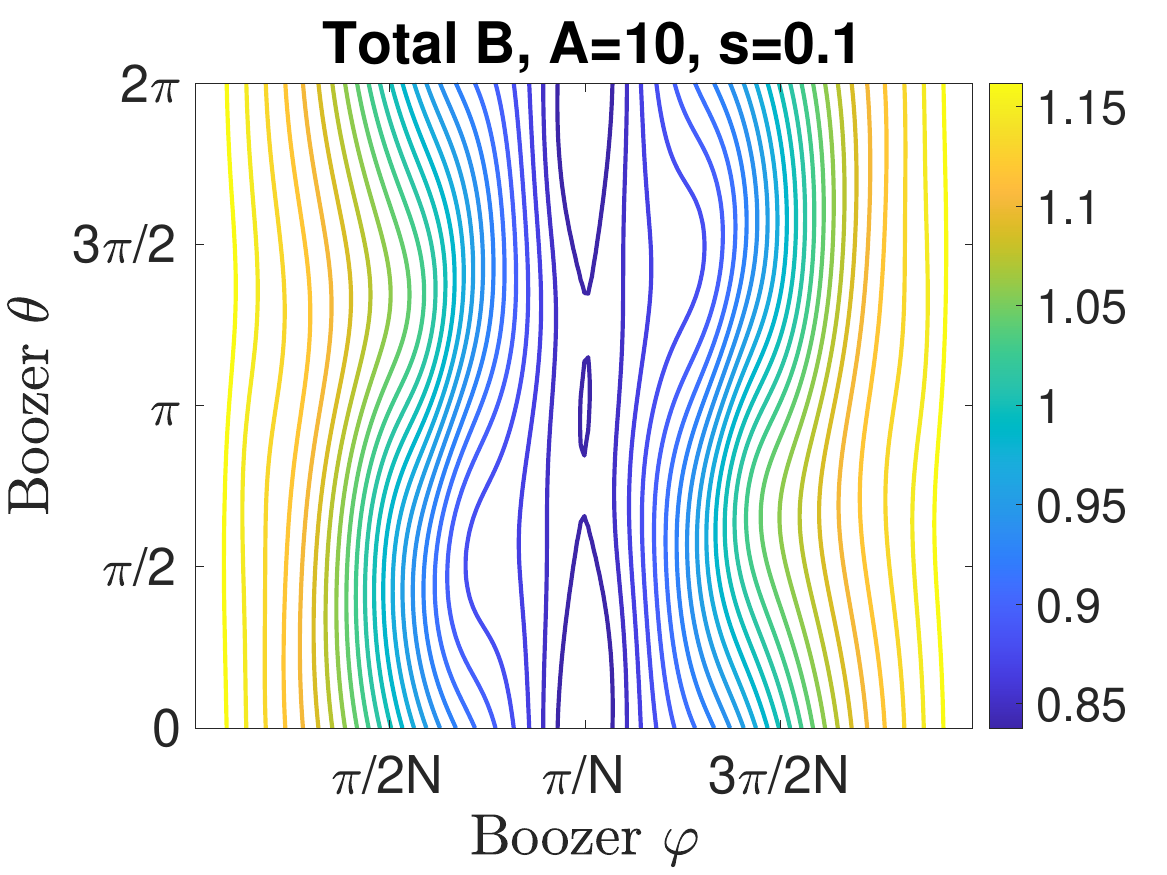}
\includegraphics[width=0.32\textwidth]{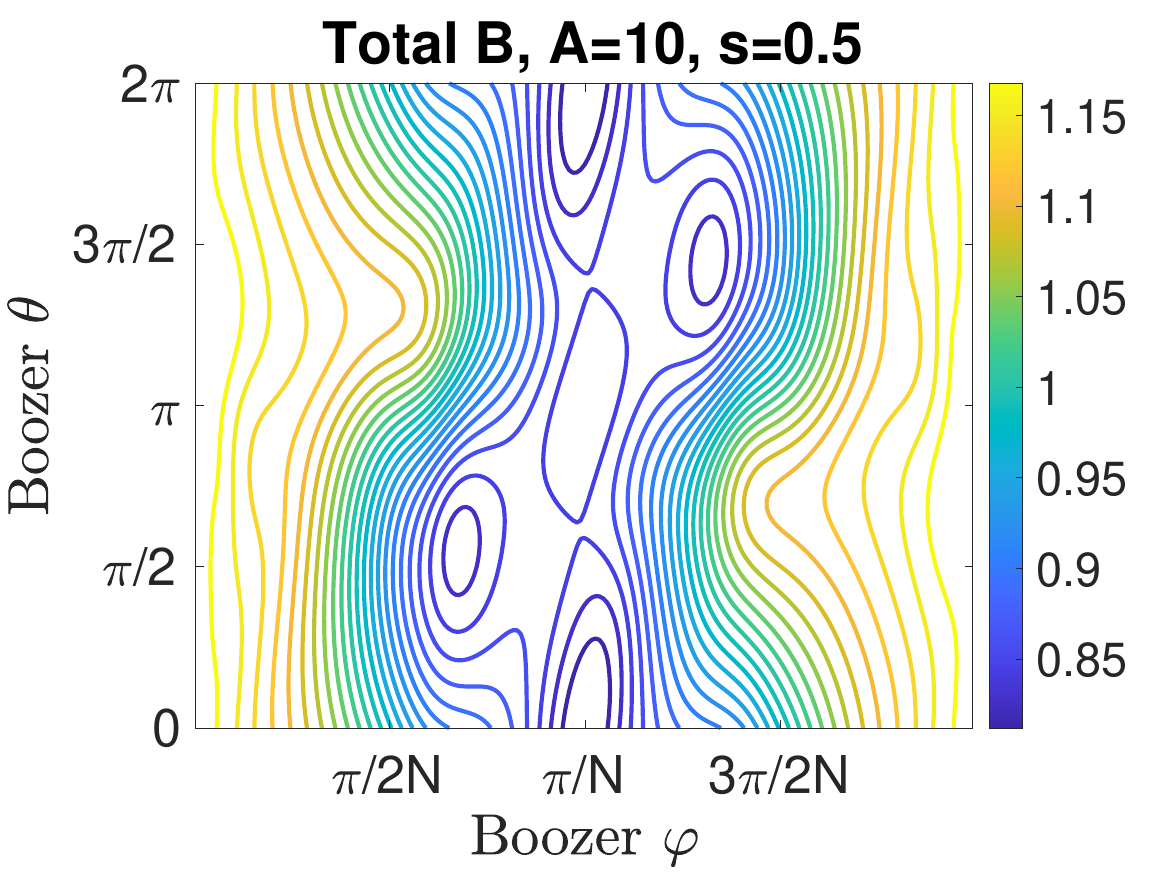}
\includegraphics[width=0.32\textwidth]{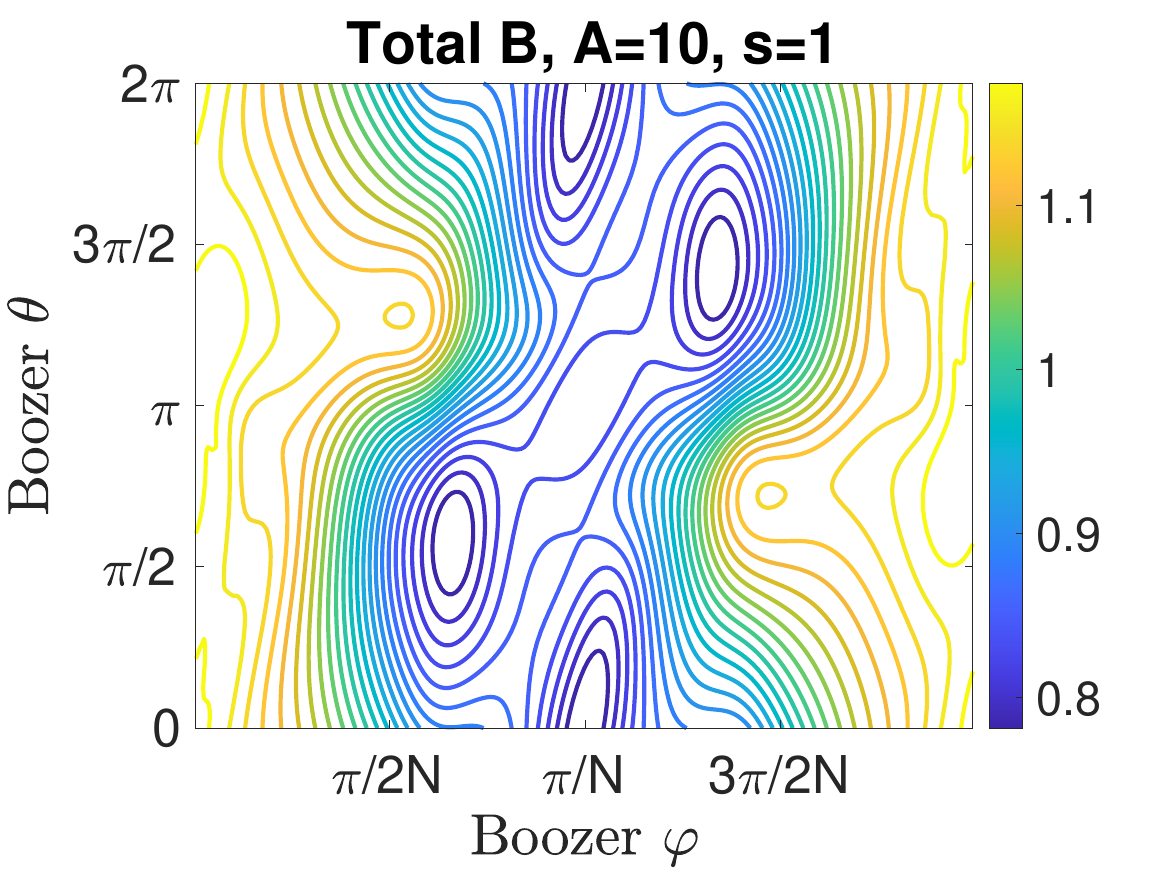} 
\caption{Contours of magnetic field intensity for the configuration in section \ref{sec:2FP_configuration} at $s{=}0.1$ (left), $s{=}0.5$ (center) and $s{=}1$ (right). The poloidally closed contours of $B$, characteristic of QI configurations, degrades with the distance from the axis.}
\label{fig:2FP_Contours}
\end{figure}

The $\alpha$ function is defined as in equation \ref{eq:alpha_final}, with $k=2$ chosen for this configuration. A near-axis solution is constructed using the aforementioned parameters at an aspect ratio $A=\sqrt{\frac{\bar{B_{0}}}{2}}\frac{R_c(0)}{\epsilon}=10$, where $\bar{B_0}$ is the average value of $B_0$. The boundary is converted to cylindrical coordinates using the method describe in section 4.1 of \citep{landreman2019direct}, which results in elliptical cross-sections as shown in figure \ref{fig:2FP_cross_sections_elongation}. This plasma boundary is then used to calculate an MHD equilibrium using the VMEC code. The intensity of the magnetic field on the last-closed flux-surface is shown in figure \ref{fig:2FP_Boundary}, which shows, even at this small aspect ratio, the behaviour expected for a quasi-isodynamic configuration, with poloidally closing contours of $|B|$ can be observed. These contours in Boozer coordinates are shown in figure \ref{fig:2FP_Contours}. As expected, the contours look "more QI" the closer we look to the axis, in the sense of having a greater degree of poloidal closure, as well as fewer island defects.   

\begin{figure}
\centering
  \includegraphics[width=0.45\textwidth]{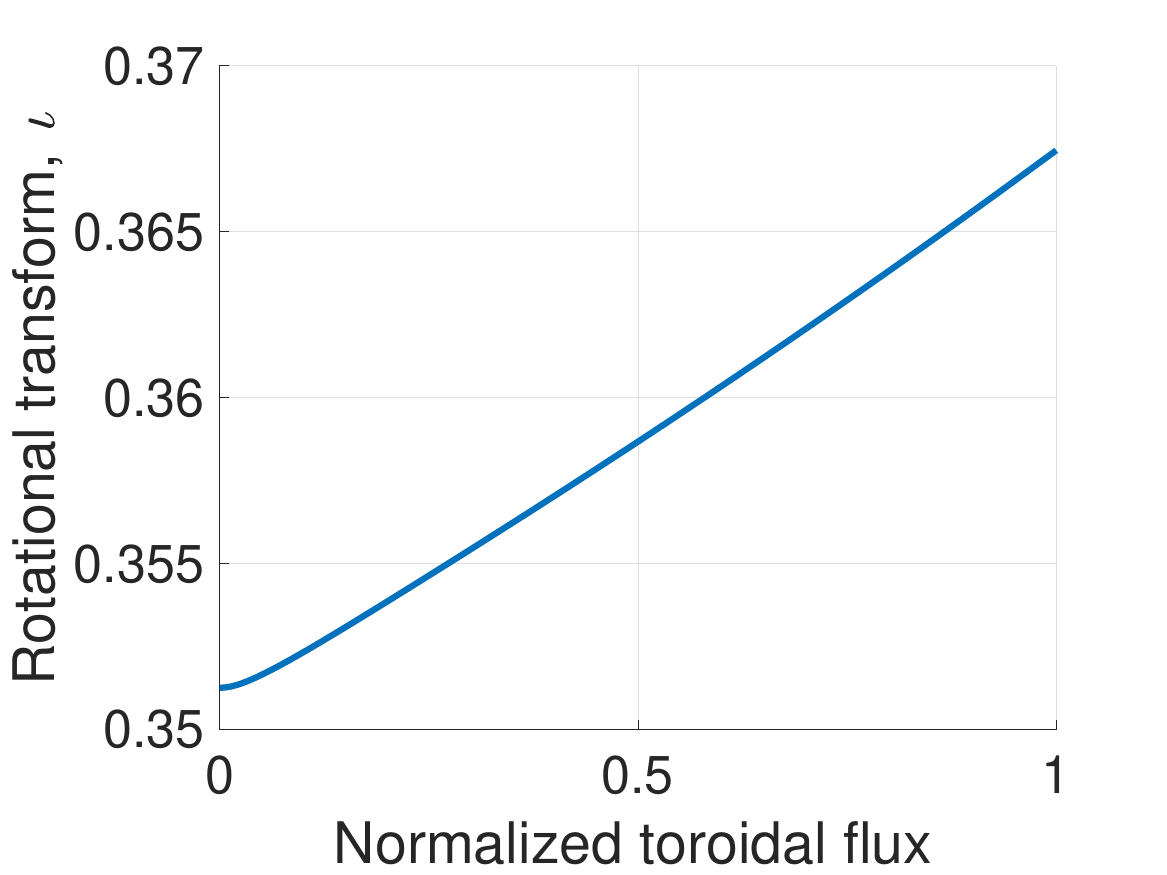}
\caption{ Rotational transform profile for the 2 field-period half-helicity configuration.}
\label{fig:2FP_iota}
\end{figure}

The resulting configuration has maximum elongation under $5.5$ and a dependence on $\phi$ as shown in figure \ref{fig:2FP_cross_sections_elongation}. Elongation here is taken at $\phi$=constant in real space, and calculated as the ratio between the maximum and minimum distances from the axis on the resulting cross-section. One of the interesting properties of half-helicity configurations is the higher values of rotational transform they can achieve compared to zero-helicity cases due to the fact that the axis curves have higher values of integrated torsion. Here, a rotational transform spanning from $\iota = 0.351-0.367$ is found for the VMEC equilibria as seen in figure \ref{fig:2FP_iota}. 

The effective ripple, $\epsilon_{\mathrm{eff}}$, is a proxy often used to estimate neoclassical transport in the $1/\nu$ regime in stellarators. We have calculated it as described in \citep{drevlak2003effective} for 16 radial points. A maximum value of $\epsilon_{\mathrm{eff}}=1\%$ is found, as shown in figure \ref{fig:eps_eff}, the values of the effective ripple are similar to those of the standard configuration of W7-X, with the important distinction that this configuration consists entirely of elliptical cross-sections and no boundary optimization was required. Half-helicity configurations seem to provide the high rotational transform benefits from $m=1$ configurations while preserving the low values of neoclassical transport found in $m=0$ equilibria. In the next section we will show we can take advantage of this property to construct  $N{>}3$ equilibria with low effective ripple.   

\section{The half-helicity space}


Now that we have adjusted the near-axis method to deal with the possibility of half-helicity axes, we proceed to identify in which region of the space of solutions we can find such configurations. Let us analyze the space of 5 field-period configurations using the same procedure as that in section \ref{Helicity}. 

Once again, we hold all parameters fixed, apart from those of the magnetic axis, and we parametrize the axis in cylindrical coordinates as

 \begin{equation}
     R(\phi) = 1 - \frac{1}{101} \cos (5\phi),
 \end{equation}
  \begin{equation}
     z(\phi) = z_s(1)\sin(5\phi) + z_s(2)\sin(10\phi).
 \end{equation}

\begin{figure}
\centering
  \includegraphics[width=0.99\textwidth]{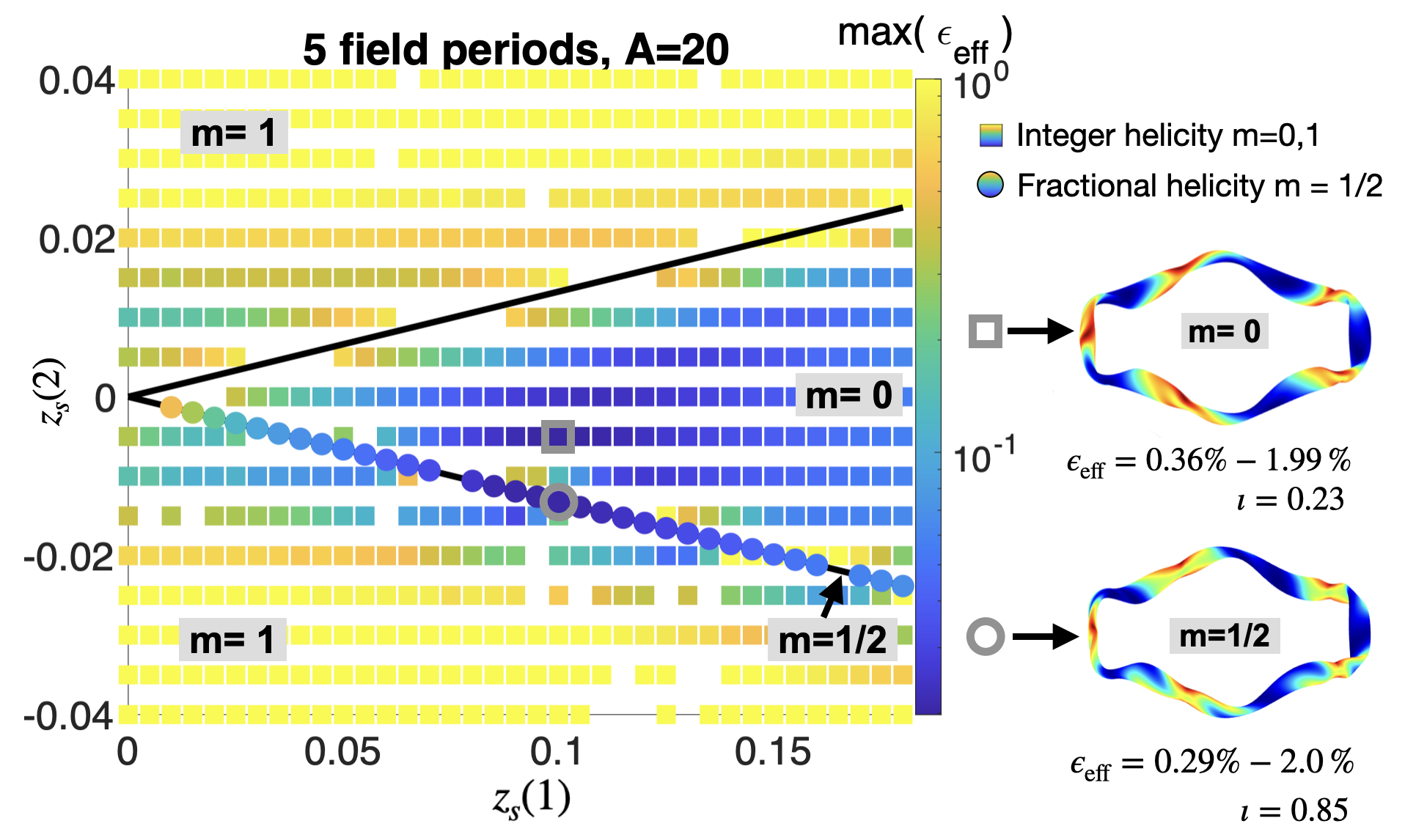}
\caption{Space of 5 field-period near-axis configurations. Squares correspond to integer helicity values (m=0,1) and circles to half-helicity (m=1/2). Colors indicate the maximum effective ripple. An example of each type of helicity is shown on the right. Note that the half-helicity configurations attain similar values of \epsE{} as integer configurations but have higher values of rotational transform.}
\label{fig:epsEff_5FP}
\end{figure}

The aspect ratio is chosen as $A=20$, the function $\alpha(\varphi)$ is described by equation \ref{eq:alpha_final} with $k=2$, the magnetic field on axis is $B_0 = 1+0.15\cos (\varphi)$ and $d = 0.73 \kappa^s$. Each color-coded square in figure \ref{fig:epsEff_5FP} is an integer-helicity axis configuration with values of $z_s(1)$ ranging from [0, 0.18] and $z_s(2)$ from  [-0.425, 0.425]  constructed using the near-axis method described in section \ref{NAE}.

  We also include half-helicity axes for which $R_c(3)$ and $z_s(3)$ in expressions \ref{eq:Axis_HalfHelicity_Conditions} are zero. Thirty seven values of $z_s(1)$ are considered, ranging from [0,0.18]. In order to fulfill the half-helicity constraints on the axis $z_s(2)$ is calculated following expression \ref{eq:Axis_HalfHelicity_Conditions}
  \begin{equation*}
      z_s(2) = \frac{2+N^2}{4+8N^2} z_1
  \end{equation*}

The configurations for these axes are constructed as in the previous section: the only difference with the integer helicity cases is the need to set $m=1/2$ when calculating the function $\alpha(\varphi)$. All other initial parameters are as described earlier in this section. Given how $z_s(1)$ depends on the chosen values for $z_s(2)$ when restricting ourselves to two parameter curves, the half-helicity space has one dimension less than the equivalent integer-helicity space. This is always the case, independently of the number of Fourier coefficients included in the axis representation.
  
 In figure \ref{fig:epsEff_5FP}, the circles correspond to the half-helicity configurations and are distributed on the line where the conditions for second order zeros of curvature at the maxima of $B_0$ are fulfilled. The line in the upper quadrant is for those axes in which $\kappa$ is zero at second order at the minima of $B_0$, these are not consistent with near-axis QI equilibria, hence the lack of configurations in said region.   
 
A configuration from the zero-helicity section is shown above a half-helicity configuration in figure \ref{fig:epsEff_5FP}. Both are the configurations with lowest \epsE{} in the boundary, in each space. We can see both have similar boundary effective ripple values (1.99\% for $m{=}0$, and 2\% for $m{=}1/2$) but the rotational transform is substantially larger for the half-helicity case ($\iota_{m=0}{=}0.23$ and $\iota_{m=1/2}{=}0.85$). This behaviour is consistently observed between all configurations in this space. The half-helicity space, apart from its lower dimensionality, also has higher values of rotational transform, closer to the per-field-period values encountered in traditional optimized stellarators. Higher values of rotational transform are desirable as they are related to improved confinement \citep{ascasibar2008effect} and given the dependence of ideal MHD $\beta$-limit on $\iotaslash$. \citep{miyamoto2005plasma, loizu2017equilibrium}


\section{5 Field-period configuration}\label{5FieldPeriods}
 
Up until now it had proven difficult to find configurations with integer helicity, high numbers of field periods, and reasonably low aspect ratio, which are accurate to first order, using VMEC as described previously. This is true when attempting to optimize within the space of near-axis solutions. We now show that this is not the case when searching in the lower-dimensional space of half-helicity magnetic axes in the near-axis expansion. 

A magnetic axis shape is chosen by varying the Fourier coefficients $R_c(3), z_s(2)$ and $z_s(3)$ of expression (\ref{eq:Axis_Truncated_FourierRepresentation}) and imposing the half-helicity conditions in (\ref{eq:Axis_HalfHelicity_Conditions}). An axis is selected such that low values of effective ripple are achieved in the boundary, the one used for the construction of the following example is described by

\begin{equation}\label{eq:R0_5FP}
    R = 1 + \tfrac{0.2260}{26}\cos{(5\phi)} - \tfrac{1}{101} \cos{(10\phi)} - 1\times10^{-3} \cos{(15\phi)} ,
\end{equation}
\begin{equation}\label{eq:Z0_5FP}
    z =  -\tfrac{3.1986}{27} \sin{(5\phi)} +7\times10^{-3} \sin{(10\phi)} + 2.6\times10^{-3}\sin{(15\phi)}.
\end{equation}

\begin{figure}
\centering
  \includegraphics[width=0.5\textwidth]{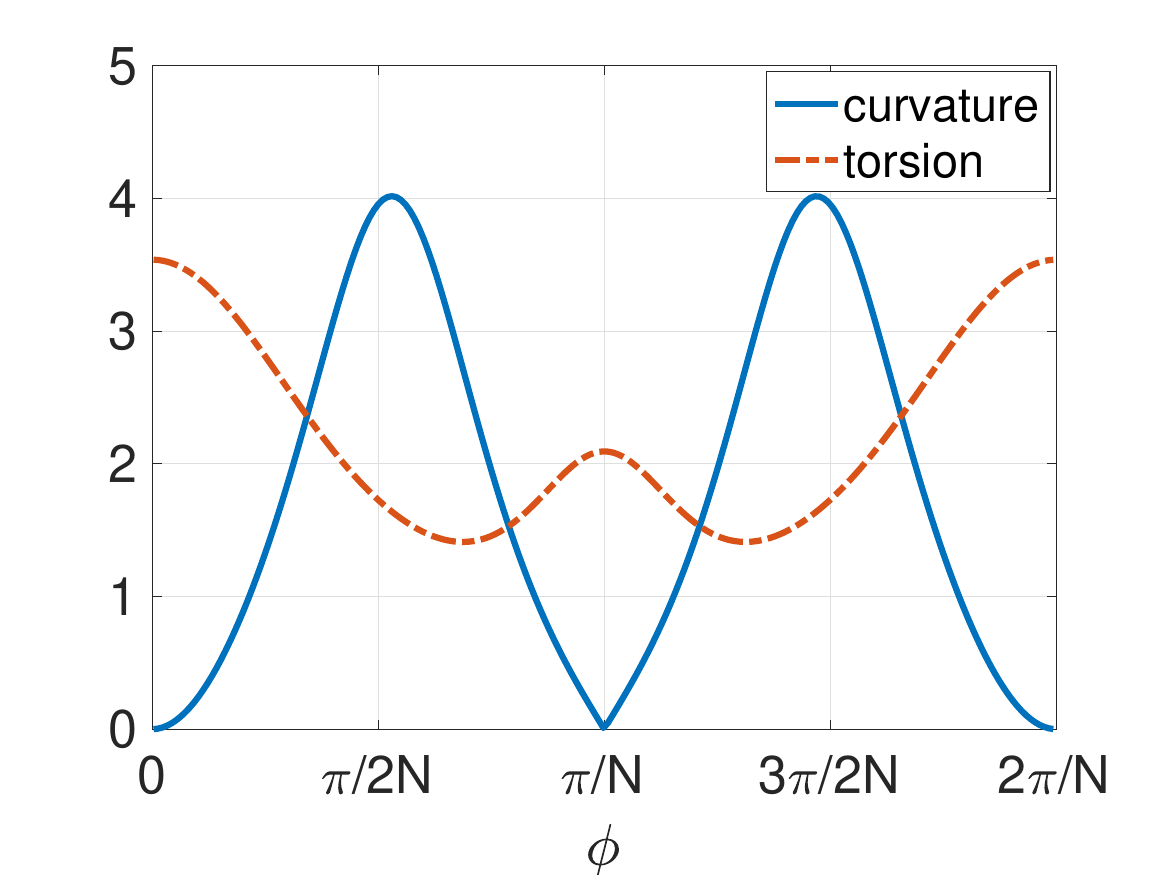}
\caption{Curvature (solid) and torsion (dashed) profiles of the 5 field-period magnetic axis described by equations \ref{eq:R0_5FP} and \ref{eq:Z0_5FP}. }
\label{fig:5FP_kappaTau}
\end{figure}

The curvature and torsion per-field period are shown in figure \ref{fig:5FP_kappaTau}. For this high number of field periods, the magnetic field on axis and the function $d(\varphi)$ need to be carefully chosen to reduce $\epsilon_{\mathrm{eff}}$ to acceptable levels. These quantities are parametrized as

\begin{equation}
   B_{0} (\varphi) = 1 + B0_1 \cos{(5\varphi)} + B0_2 \cos{(10\varphi)}, 
\end{equation}
\begin{equation}\label{eq:d_param}
     d(\varphi) = \sqrt{d_{\kappa}/B_{0}(\varphi))}\kappa^s + d_{\kappa c} \kappa^s  \cos{(5\phi)} + d_{\kappa s} \kappa  \sin{(5\phi)}.
\end{equation}

\begin{figure}
\centering
  \includegraphics[width=0.49\textwidth]{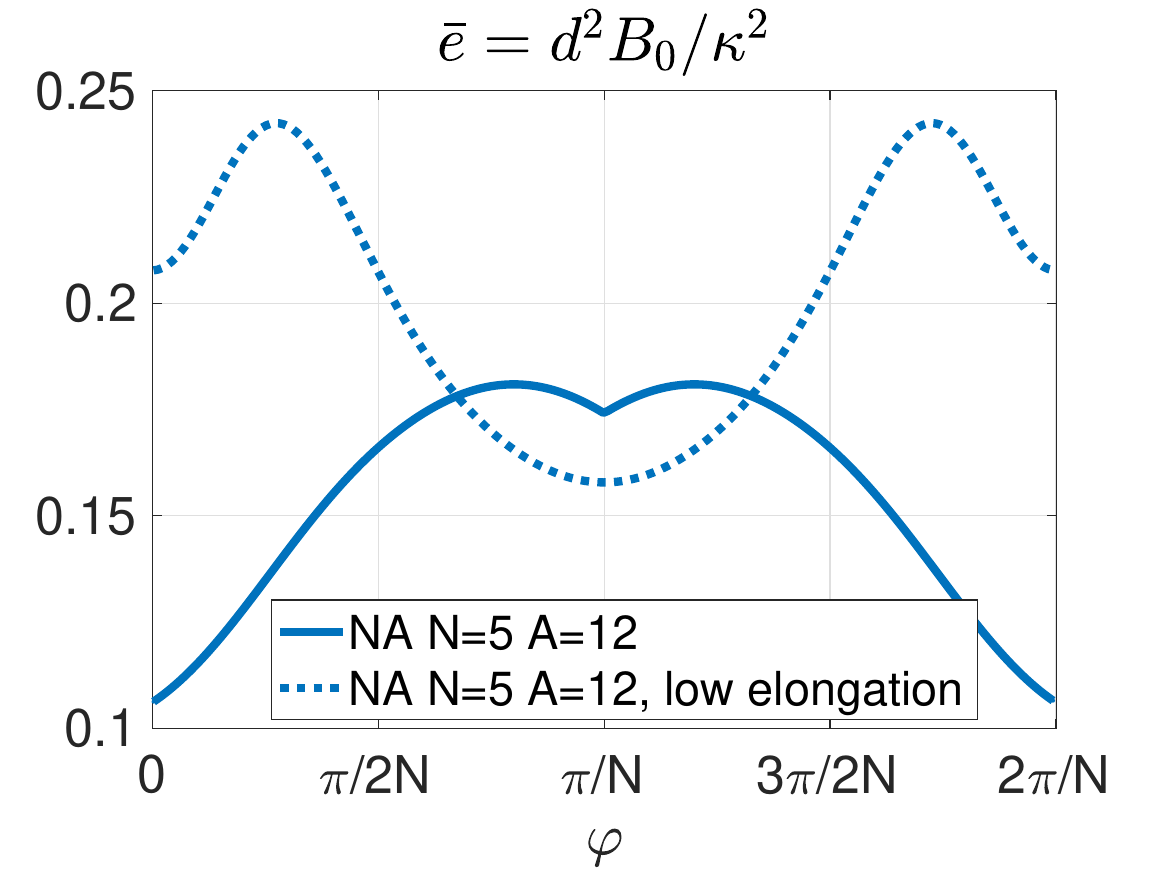}
  \includegraphics[width=0.49\textwidth]{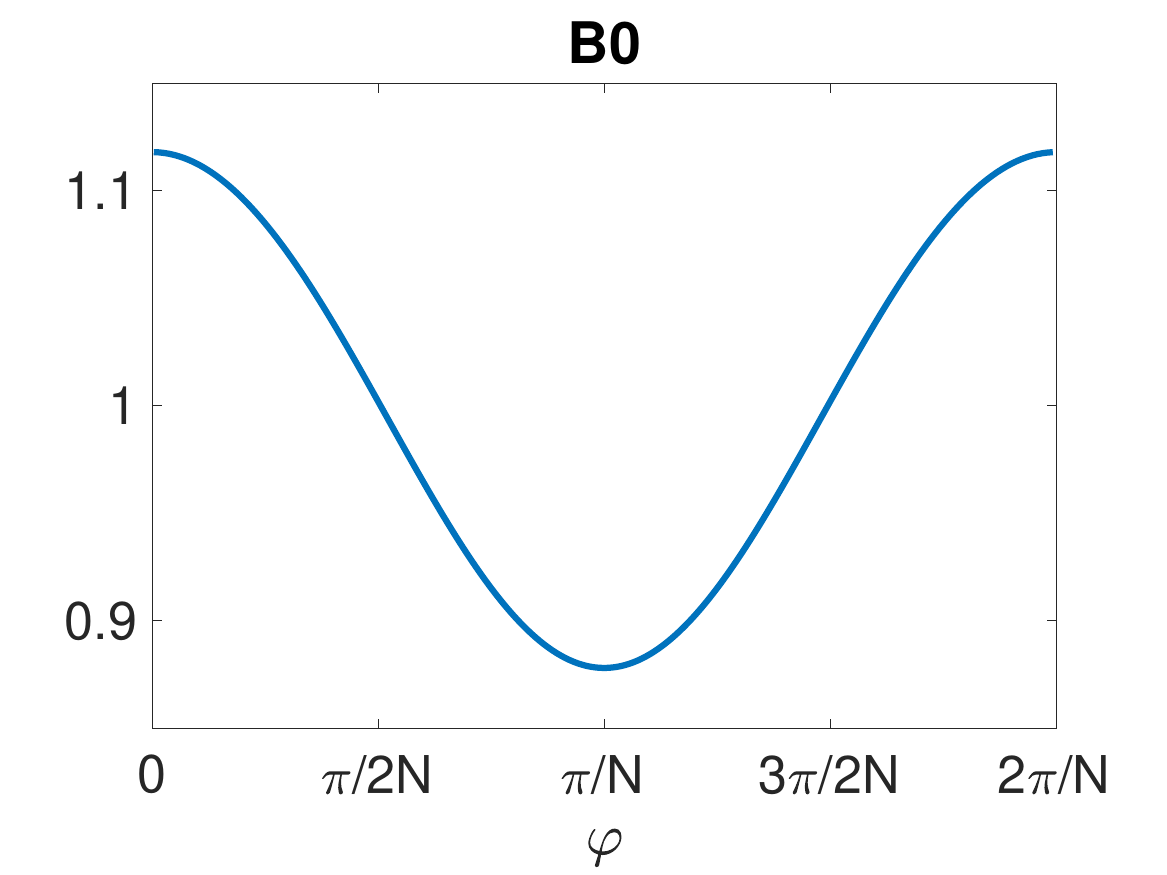}
\caption{(left) $\bar{e}$ profiles for the two 5 field-period configurations presented,  dotted line correspond to the configuration with parameter $d(\varphi)$ optimized for low elongation. (right) Magnetic field on axis $B_0(\varphi)$ used for the construction of these configurations. }
\label{fig:5FP_dBar_B0}
\end{figure}

\begin{figure}
\centering
  \includegraphics[width=0.49\textwidth]{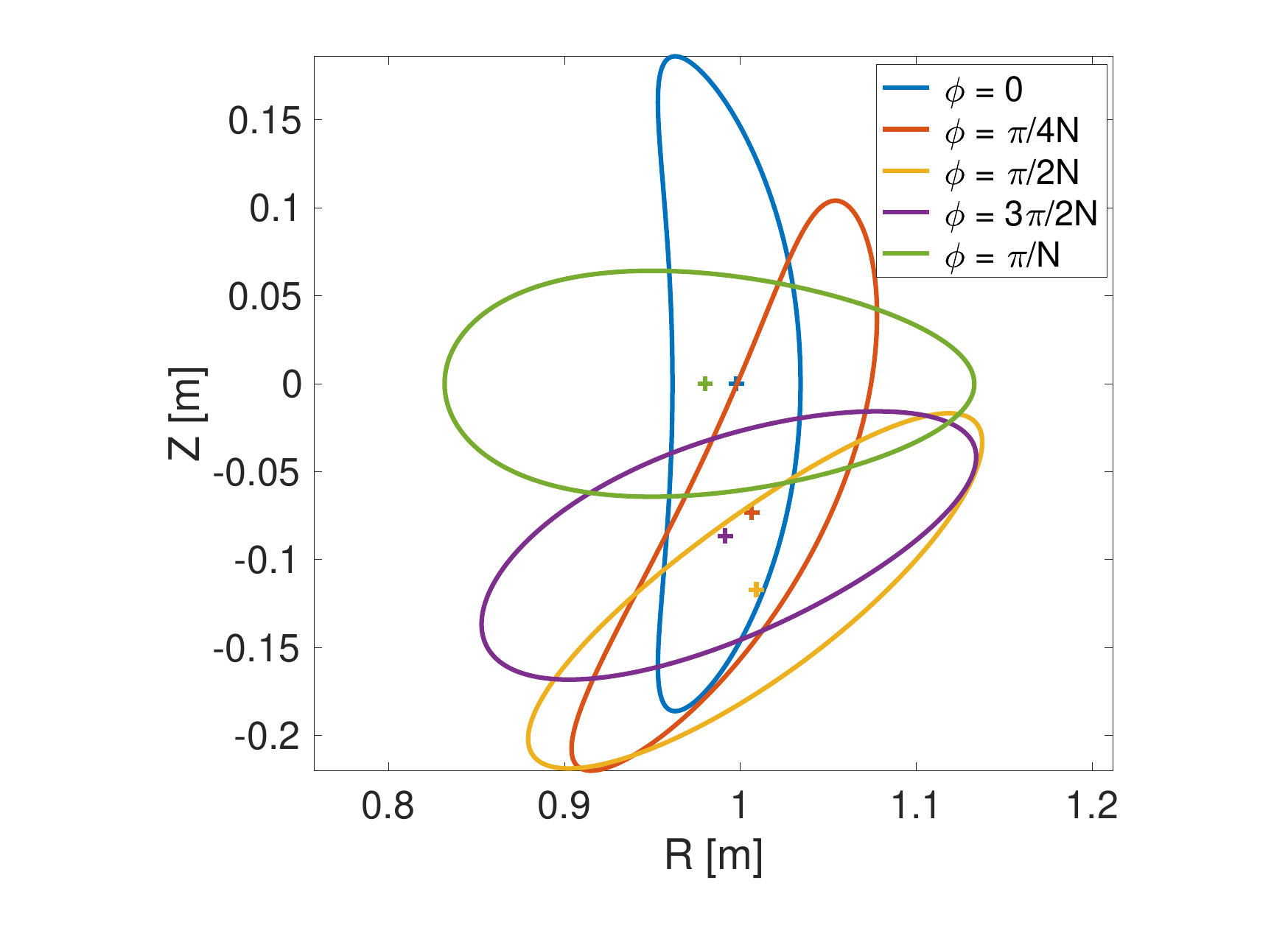}
   \includegraphics[width=0.47\textwidth]{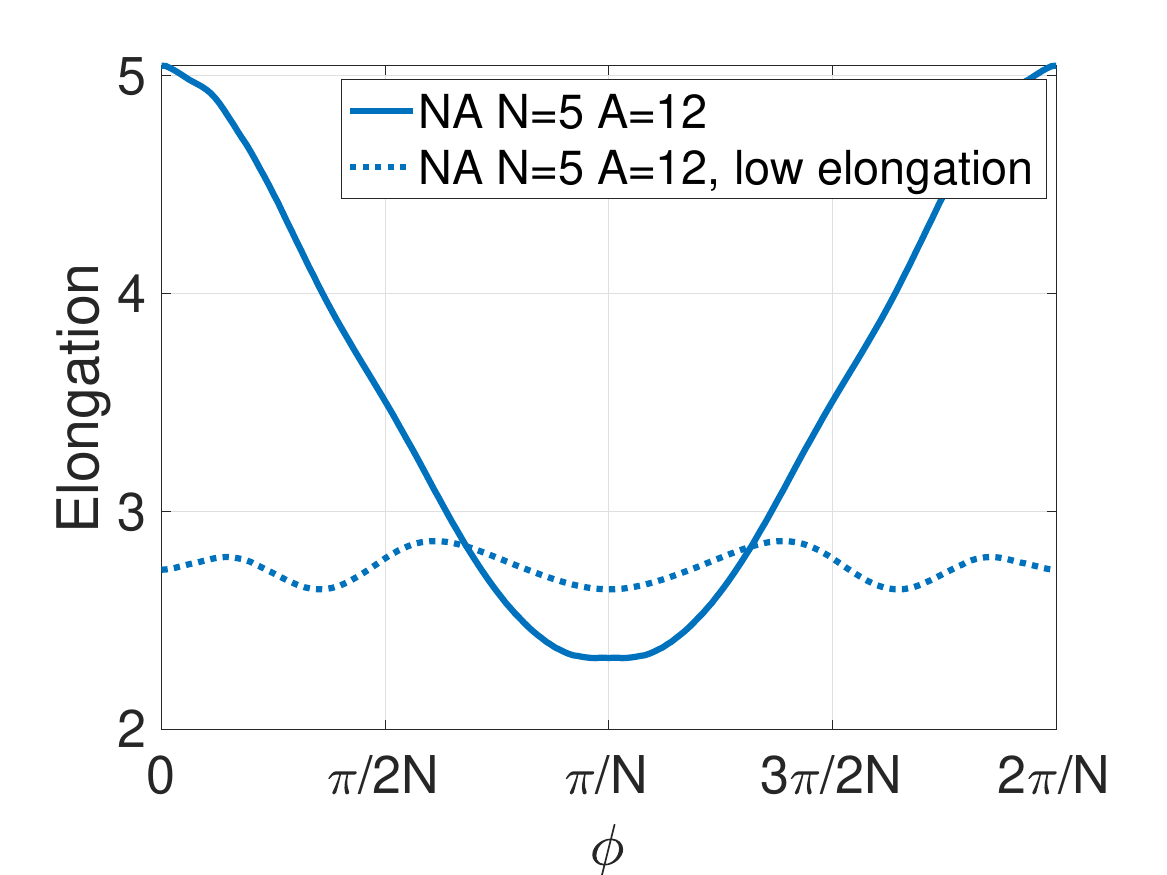} 
\caption{(left) Cross-sections of the 5 field-period configuration shown in figure \ref{fig:5FP_modB_Boozer}. (right) Elongation profiles for the two 5 field-period configurations shown in section \ref{5FieldPeriods}, dotted line correspond to the configuration with parameter $d(\varphi)$ optimized for low elongation.}
\label{fig:5FP_crossSections_elongation}
\end{figure}

Adding more parameters to these expressions allows for a finer profile control. These parameters are varied for a set axis shape until a desired configuration is found. For this example  $ d_{\kappa}=0.28$, $ d_{\kappa c}=-0.065$, $ d_{\kappa s}=0.04$, $B0_1 = 0.12$, and $B0_2 = -0.002$, the resulting shape of $B_0(\varphi)$ and $\bar{d}(\varphi)$ are shown in figure \ref{fig:5FP_dBar_B0}, here we can observe a non-analytic behaviour present at $\varphi=\pi/N$, this arises due to the sine term in expression \ref{eq:d_param} and can be set to zero but including it resulted in lower values of $\epsilon_{\mathrm{eff}}$, and, most importantly, it does not result in any sharp behaviour on the plasma boundary. As in the previous section, the function $\alpha$ is defined through equation \ref{eq:alpha_final} with $k=2$. Following the near-axis expansion formalism, we construct a near-axis solution at a set distance from the axis, in this case $r=1/12$ corresponding to an aspect ratio $A=12$. However, using the method  of section 4.1 of \citep{landreman2019direct} was not enough to obtain equilibria with good confinement, instead the method of section 4.2 of the same publication was used. This method, albeit more complicated, results in boundary shapes identical at first order to the boundary in the $X$-$Y$ space, with the caveat that cross-sections in the $R$-$z$ plane are not elliptical anymore, as can be seen in figure \ref{fig:5FP_crossSections_elongation}. Having a boundary defined by elliptical cross-sections requires the use of less poloidal modes when using equilibrium codes like VMEC and might be useful for plasma boundary optimization, where initial points with low modes are preferred, although not necessary \citep{landreman2022mapping}. 

\begin{figure}
\centering
  \includegraphics[width=0.99\textwidth]{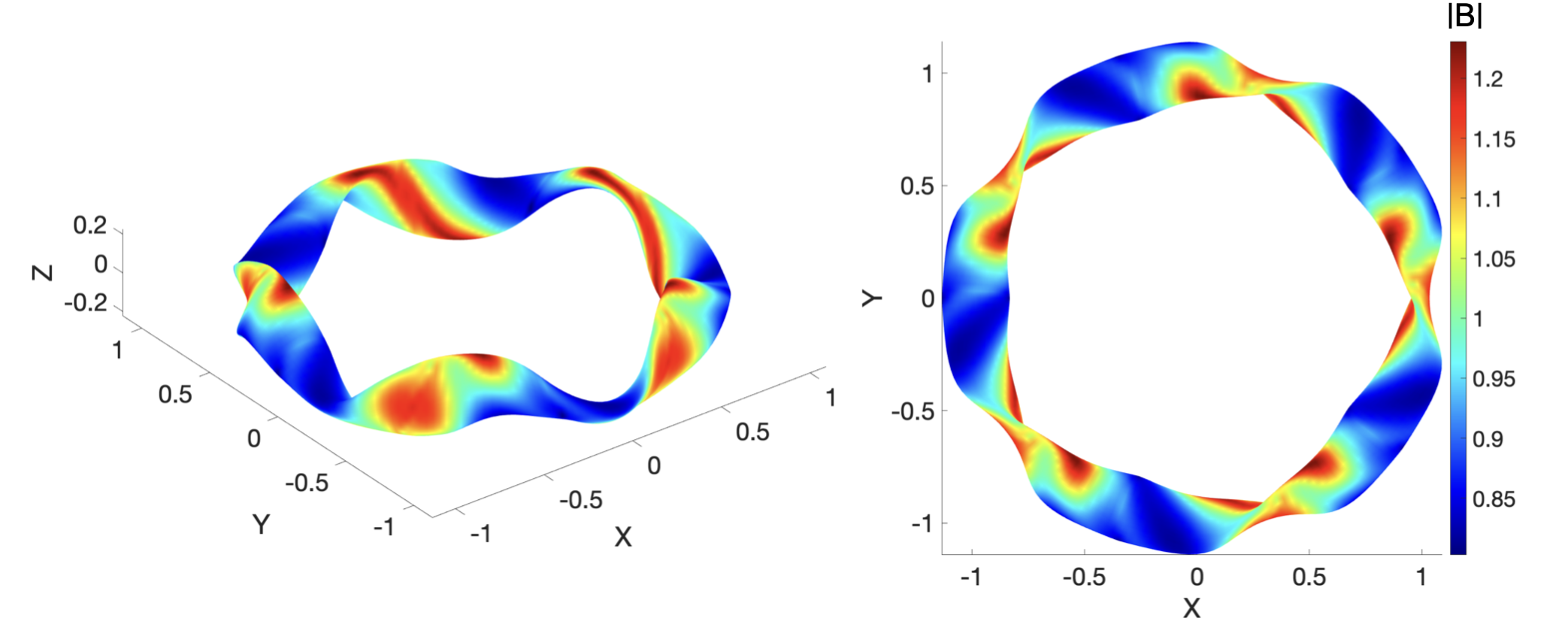}
\caption{ Intensity of the magnetic field on the boundary of the 5 field-period configuration. (left) side and (right) top view.}
\label{fig:5FP_modB_Boundary}
\end{figure}

\begin{figure}
\centering
  \includegraphics[width=0.32\textwidth]{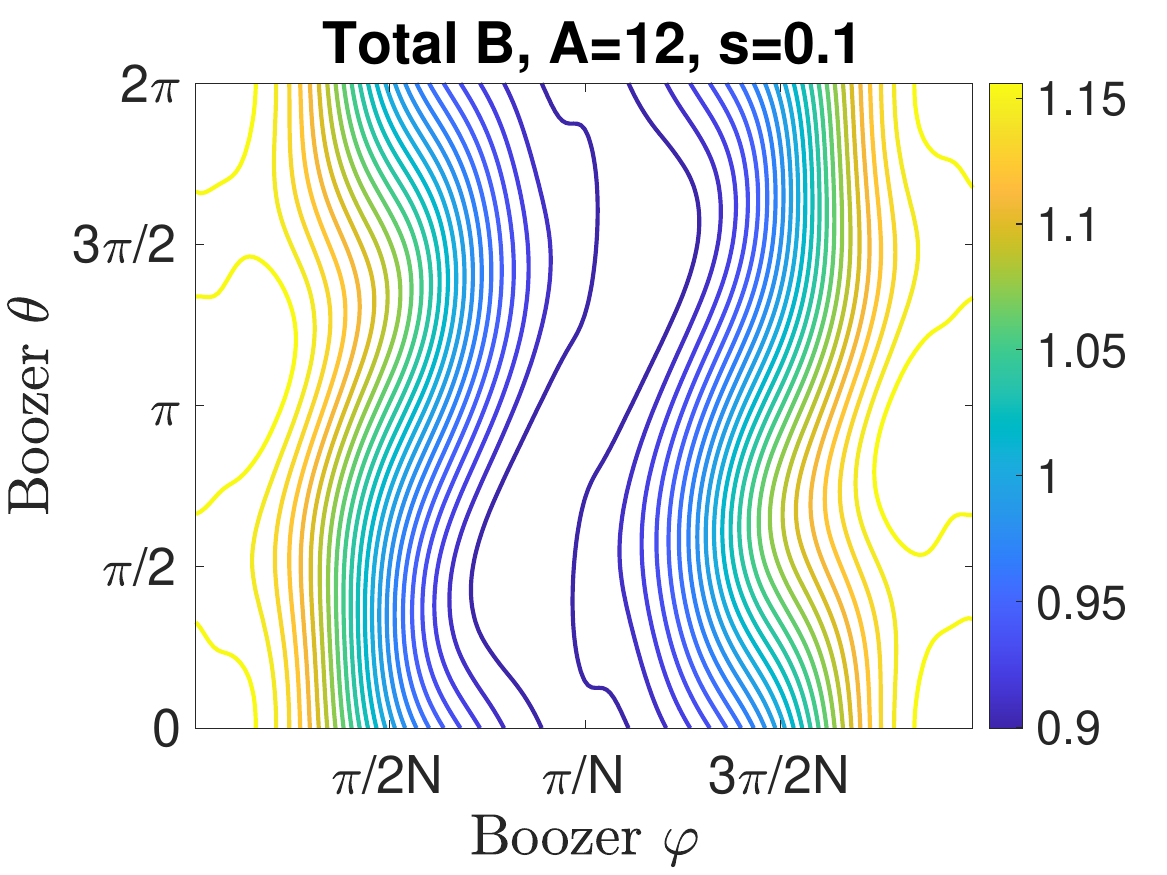}
   \includegraphics[width=0.32\textwidth]{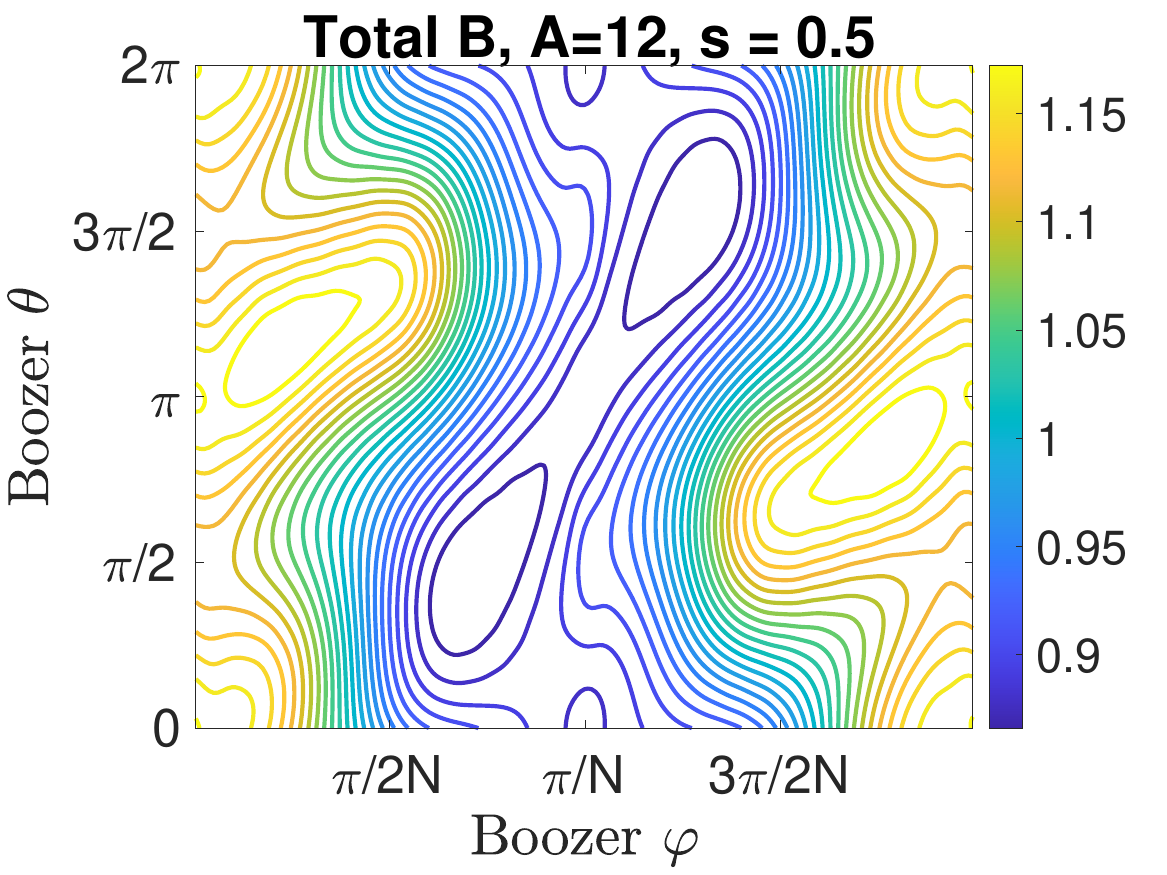}
   \includegraphics[width=0.32\textwidth]{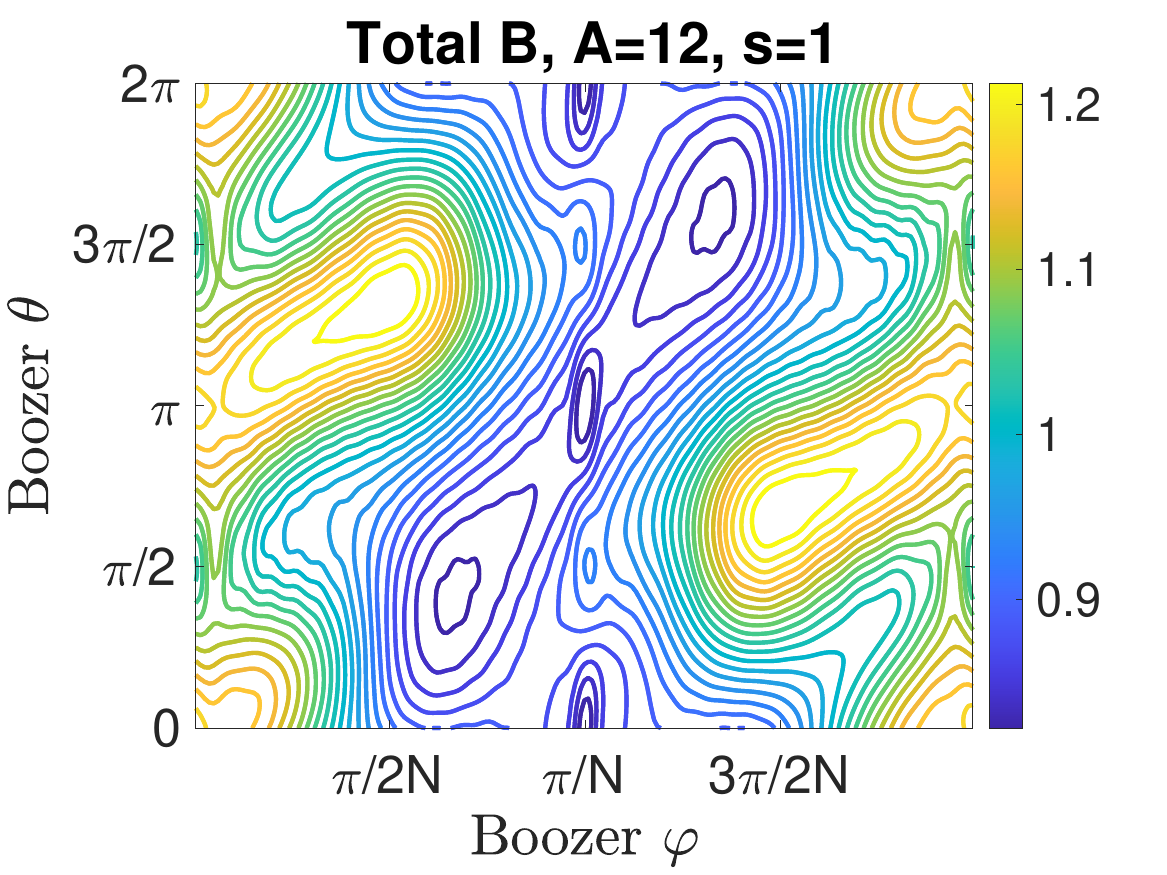}  
\caption{ Contours of magnetic field intensity for the configuration in section \ref{5FieldPeriods} at $s{=}0.1$ (left), $s{=}0.5$ (center) and $s{=}1$ (right). The poloidally closed contours of $B$, characteristic of QI configurations, degrades with the distance from the axis.}
\label{fig:5FP_modB_Boozer}
\end{figure}

\begin{figure}
\centering
  \includegraphics[width=0.99\textwidth]{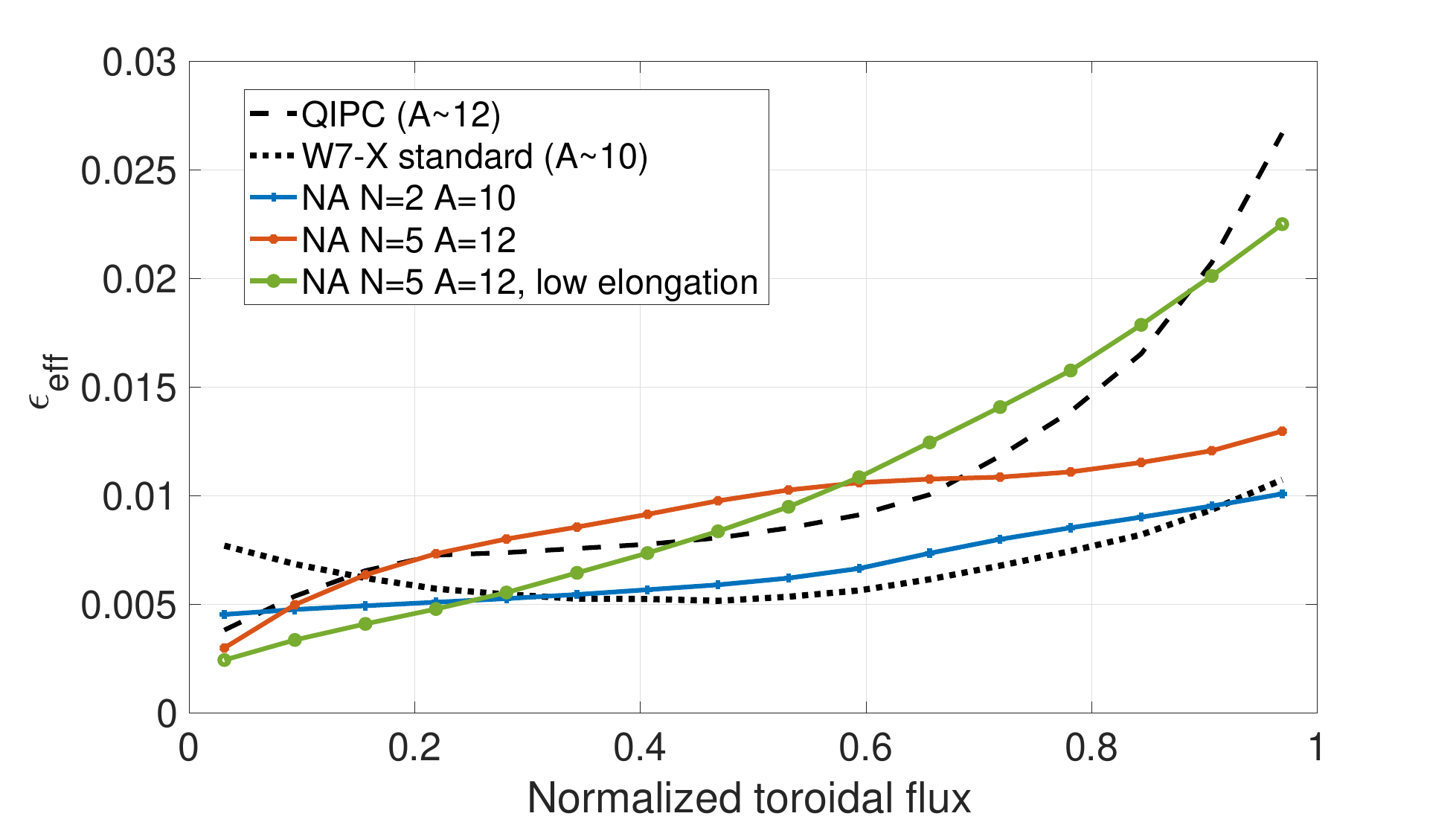}
\caption{Effective helical ripple, \epsE{} for the three configurations shown in this work (solid lines), and two quasi-isodynamic optimized configurations; W7-X (dotted line) and QIPC (dashed line).}
\label{fig:eps_eff}
\end{figure}

The intensity of the magnetic field on the boundary, as obtained from VMEC is shown in figure \ref{fig:5FP_modB_Boundary}. Despite the contours of $|B|$ looking far from the expected behaviour for a quasi-isodynamic field, this configuration has good neoclassical confinement properties, as it is evident from its effective ripple profile in figure \ref{fig:eps_eff}, with an \epsE{}${\approx}1.3\%$ it is  comparable to that of W7-X and QIPC without the need of performing a boundary optimization. As is typical for near-axis configurations, the contours of $|B|$ resemble more straight poloidally closed contours the closer we look to the axis, as seen in figure \ref{fig:5FP_modB_Boozer}. 

The elongation of the plasma boundary cross-sections in the cylindrical angle $\phi$ is shown in figure \ref{fig:5FP_crossSections_elongation}, a maximum value of $e=5.05$ is found. We can observe a clear difference in the elongation profile of this configuration and that of the 2 field-period example shown in the previous section \ref{fig:2FP_cross_sections_elongation}. 

\begin{figure}
\centering
  \includegraphics[width=0.49\textwidth]{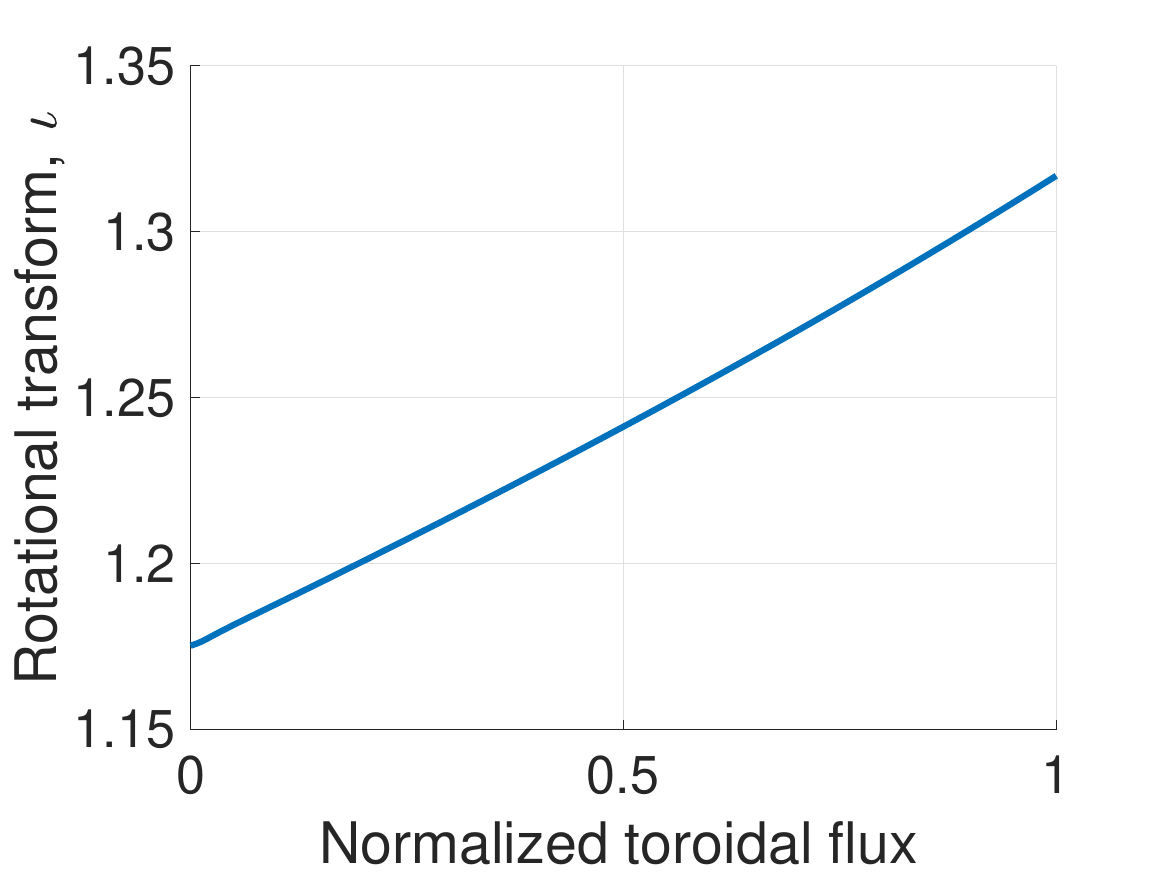}
\caption{Rotational transform profile for the 5 field-period configuration of section \ref{5FieldPeriods}. }
\label{fig:5FP_iota}
\end{figure} 

\subsection{Optimizing d for low elongation}\label{5FP_lowElongation}

The choice of the function $d(\varphi)$ plays an important role in the resulting elongation of the plasma boundary and needs to be chosen carefully, as most values will result in impractically large elongations. This function can be tuned to obtain a desired elongation profile, although the dependence on this and other quantities is not straightforward, as is clear from equation (5.1) of \citep{camacho2022direct}    

\begin{equation*}
     e(\varphi) = \frac{B_{0}\bar{d}^{2}}{4} + \frac{1}{B_{0}\bar{d}^2}(1+\sigma^2) + \sqrt{\left( \frac{B_{0}\bar{d}^{2}}{4}\right)^2 +  \frac{(\sigma^2-1)}{2} + \left(\frac{1}{B_{0}\bar{d}^2}(1+\sigma^2)\right)^{2}}.\footnote{This expression calculates elongation's dependence on the Boozer toroidal angle $\varphi$.  Geometrically, this is the elongation of the elliptical cross sections made by cutting the plasma volume perpendicular to the magnetic axis. This is not strictly the same as the profile $e(\phi)$, which is shown in figures (\ref{fig:2FP_cross_sections_elongation}) and (\ref{fig:5FP_crossSections_elongation}), where elongation is derived from cross sections formed by cuts along planes of constant $\phi$. However, the two quantities behave in a qualitatively similar manner. The VMEC-elongation, $e(\phi)$, usually larger, can be used for direct comparison with existing (VMEC) equilibria, hence shown in this work.}
\end{equation*}
 
\begin{figure}
\centering
      \includegraphics[width=0.69\textwidth]{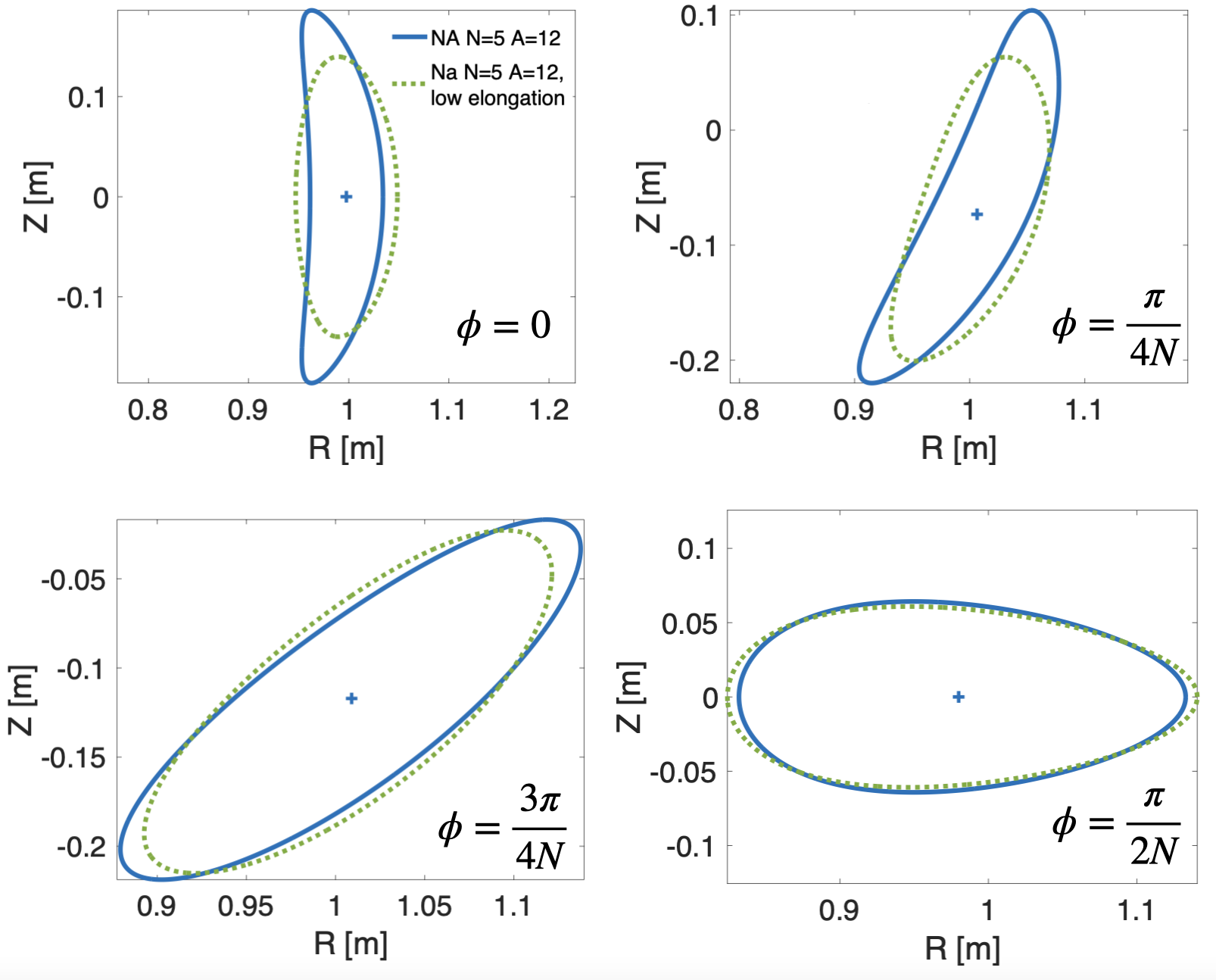}
\caption{Cross-section comparisons between the two 5 field-period configuration shown in section \ref{5FieldPeriods} at different toroidal locations. Dotted lines correspond to the configuration optimized for constant elongation.}
\label{fig:5FP_crossSections_comparison}
\end{figure}

For this reason, in the following configuration, $d(\varphi)$ is chosen through an optimization procedure to minimize elongation. All other near-axis parameters are kept the same as in the previous example. The  function $d$ is parametrized as 

\begin{equation}\label{eq:ebar}
d(\varphi) = \sqrt{d_{\kappa}/B_{0}(\varphi))}\kappa^s + \sum_{i=1}^4 d_{\kappa c}(i)\kappa^s  \cos{(5i\phi)}, 
\end{equation}
and the coefficients $d_{\kappa c}(i)$ are optimized using an off-the-shelf optimization method\footnote{The optimization method used is the Nelder-Mead simplex algorithm as described in \cite{lagarias1998convergence} and used in the \textit{fminsearch} MATLAB function. }  to reduce the elongation variation, \ie{} the difference between the maximum and minimum elongation, both outputs of the near-axis expansion construction. For this configuration the values resulting in the elongation shown with a dotted line in figure \ref{fig:5FP_crossSections_elongation} are
\begin{equation*}
    d_{\kappa} = 0.4, \qquad d_{\kappa c} = [0.0591,-0.0282,-0.0157,-0.0038],
\end{equation*}
corresponding to $\bar{e}(\varphi)$ shown in figure \ref{fig:5FP_dBar_B0}, we can see the variation in the elongation profile is greatly reduced compared to the previous configuration, which is also evident by comparing the toroidal cross-sections as done in figure \ref{fig:5FP_crossSections_comparison}. But even using four Fourier modes in the representation of $d(\varphi)$ is not enough to obtain a constant elongation profile, which is a particularly attractive case allowed by the NAE QI theory due to its specially smooth boundary behavior. This fact, together with the complicated dependence of elongation, and the parameters required for the equilibrium construction (see equation \ref{eq:ebar}), as well as the non-analyticity of $\bar{d}(\varphi)$ necessary to achieve good confinement in the example in section \ref{5FieldPeriods} are clear indications of the unsuitability of $d(\varphi)$ as a natural input parameter for the near-axis expansion. A possible solution to this problem is specifying the elongation function as an input parameter instead, this will be described in a future publication.

\begin{figure}
\centering
      \includegraphics[width=0.89\textwidth]{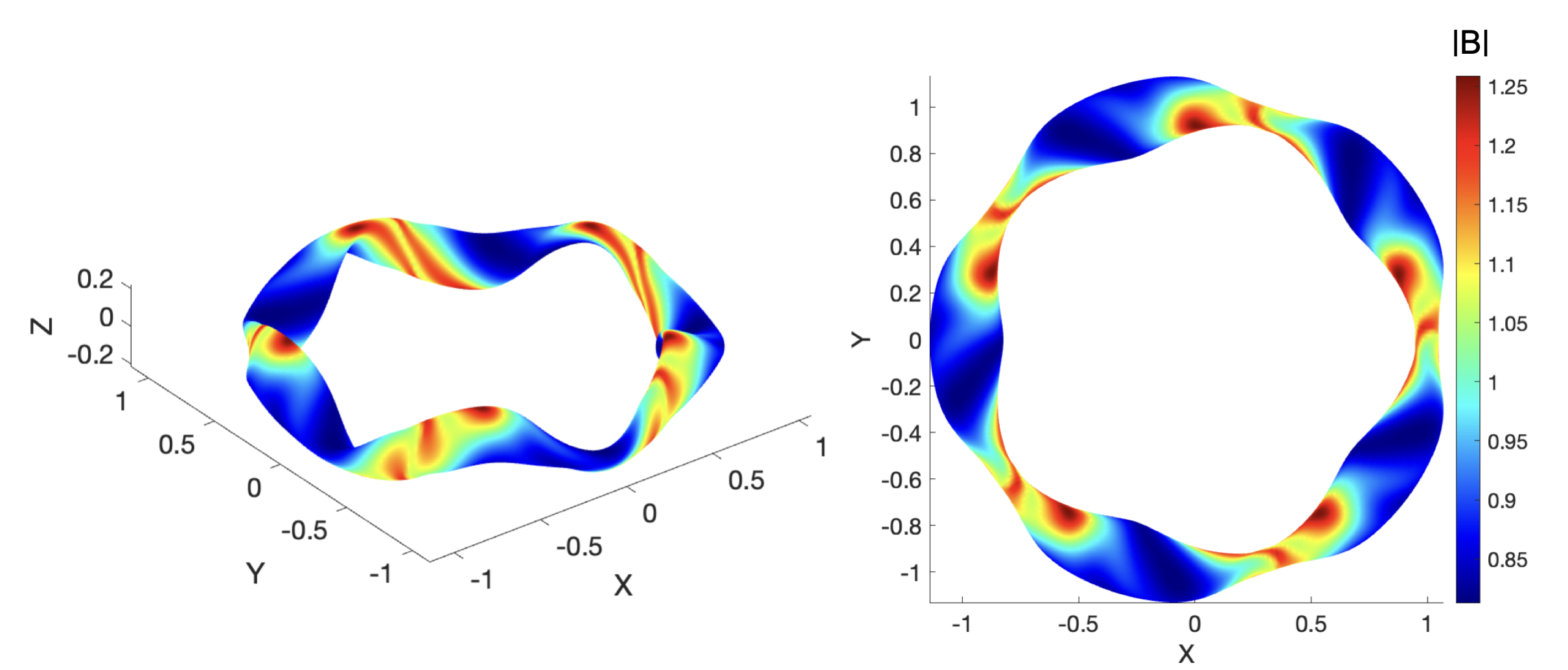}
\caption{Magnetic field intensity on the boundary for the 5 field-period low elongation configuration described in section \ref{5FP_lowElongation}. (left) Side and (right) top view.}
\label{fig:5FP_modB_Boundary_lowElongation}
\end{figure}

The intensity of the magnetic field on the boundary ($A=12$) for the configuration constructed in this section is shown in figure \ref{fig:5FP_modB_Boundary_lowElongation}, we can see the contours at maximum field do not close as expected but nonetheless $\epsilon_{\mathrm{eff}}$ remains under $1\%$ for half flux and is comparable to that of QIPC in the remaining region as shown in figure \ref{fig:eps_eff}. As expected, a reduction in elongation results in worse QI quality as discussed in \citep{camacho2022direct}.    

\begin{figure}
\centering
      \includegraphics[width=0.49\textwidth]{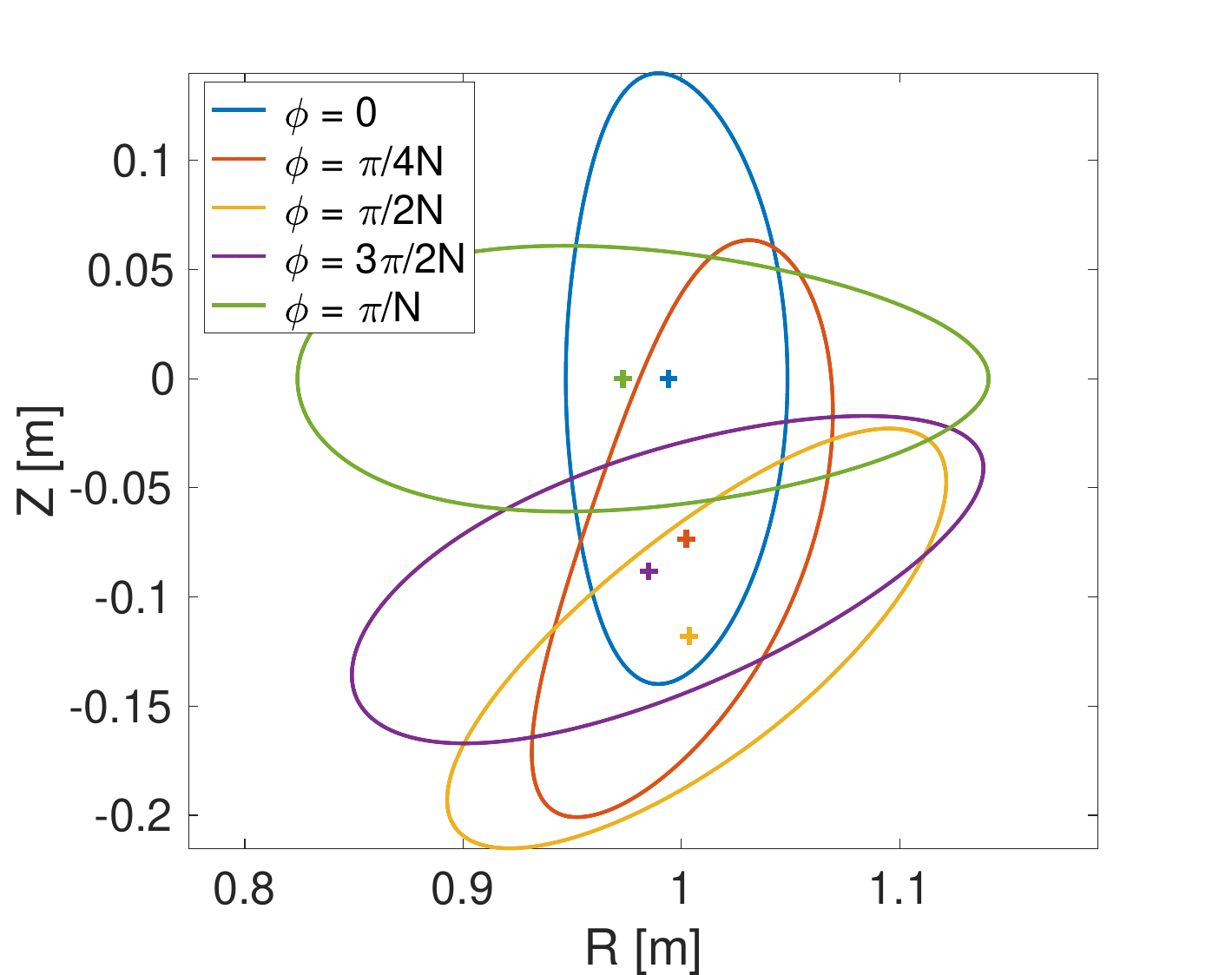}
   \includegraphics[width=0.49\textwidth]{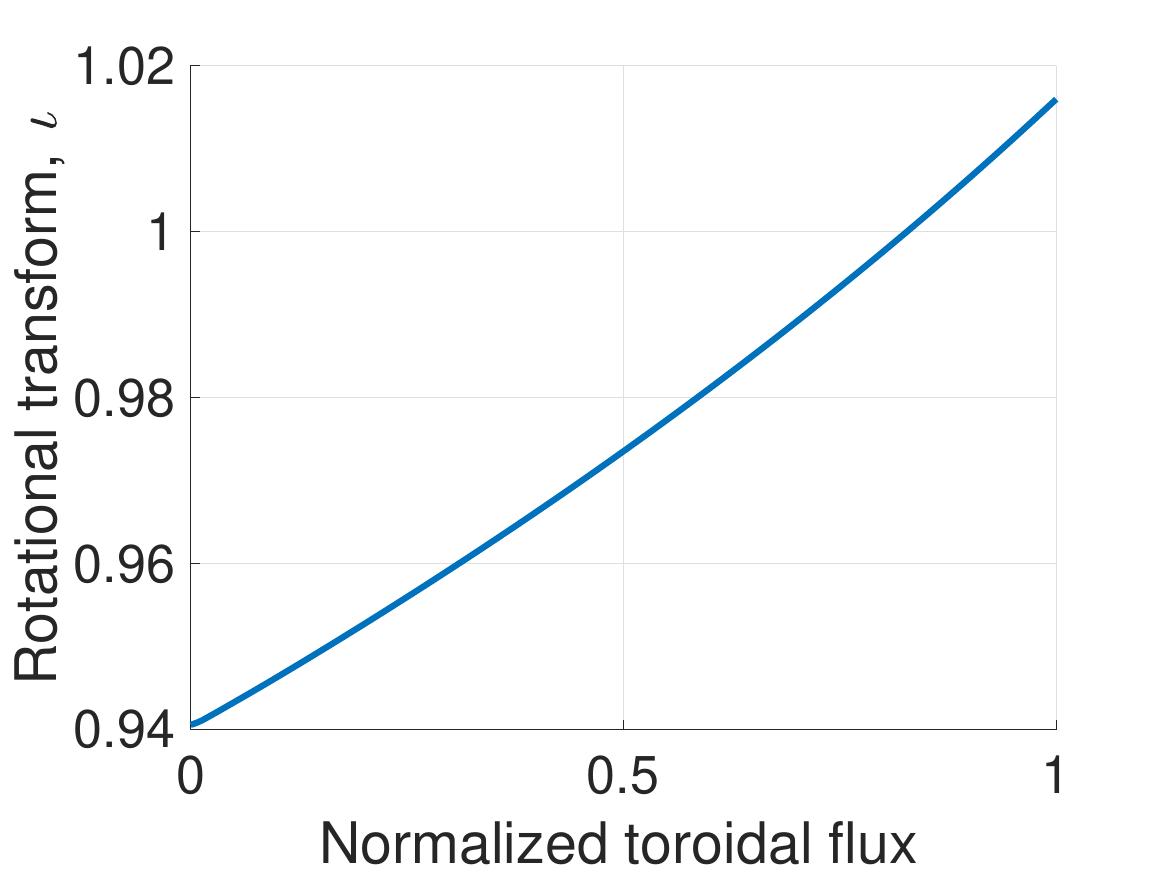} 
\caption{(left)Cross-sections of the configuration shown in figure \ref{fig:5FP_modB_Boundary_lowElongation}. (right) Rotational transform profile for the same configuration.}
\label{fig:5FP_crossSections_iota}
\end{figure} 

\section{Conclusions}

Using the near-axis expansion method, at first order, we have analyzed the impact the magnetic axis helicity has on the quality of the construction of quasi-isodynamic stellarator-symmetric configurations.

The difficulties encountered when performing stellarator boundary optimization have often been associated to the complexity of the solution space. As a first step to understand the landscape of solutions, we generated a multitude of configurations with identical NA parameters, differing only in their axis shape. This was parametrized by two terms in a Fourier representation of the axis. We show how the helicity of the magnetic axis divides the space of quasi-isodynamic near-axis configurations with clear divisions between the considered helicities, $m{=}0$ and $m{=}1$. The effective ripple was calculated for each of the constructed configurations and a region of bad quality configurations (extremely elongated or effective ripple larger than 1\%) was observed in the helicity transition region. Notably, consistently lower values of effective ripple were observed for the zero-helicity region compared to helicity one, with this behaviour being present at different number of field-periods. However, given the relation between helicity and integrated torsion, the configurations in the $\mathrm{m}{=}0$ region have low values of rotational transform. The presence of a sharp division between helicity regions would make it challenging for an optimizer to find solutions outside of its starting region.    

Stellarators such as QIPC \citep{subbotin2006integrated} that were optimized by conventional means rather than through the near-axis expansion sometimes possess a half-helicity axis, so that the normal vector performs exactly half a rotation around the axis in every field-period. 
The conditions on the coefficients of a Fourier series necessary for achieving this behaviour were calculated and correspond to the transition between integer helicity regions.  We show half-helicity axes are compatible with the near-axis formalism when modifying the $\alpha(\varphi)$ function accordingly. We present a 2-field-period configuration constructed around one such axis, at an aspect ratio $A{=}10$ the maximum \epsE{} is 1\%,  $\iotaslash{=}0.35-0.37$, and the maximum elongation remains under $e{=}5.5$. 

The case of 5 field periods was investigated, locating the half-helicity configurations within the solution space. We note that, when using a set number of Fourier axis modes, the half-helicity space has one fewer dimension compared to the integer-helicity space. This is specially relevant when a low number of Fourier modes are being used as is in this work. 

In this way, configurations with $N{=}5$ were found with at an aspect ratio $A{=}12$ and lower  effective ripple (\epsE{}${\approx}1.3\%$) than previously achieved  for near-axis QI configurations with this high number of field-periods. We also demonstrate the utility of $d(\varphi)$ in managing elongation, by showing a configuration with a low toroidal variation of this quantity. It is noted that this control over elongation, due to the freedom in $d$, is not possible for quasi-symmetric configurations, and it is planned to investigate this further in a future publication. 

The use of half-helicity axes makes it possible to construct QI stellarators with similar characteristics to those of existing traditionally-optimized QI devices, and opens the door to exploring the space of solutions in a systematic way by using near-axis configurations in the high-N region of the space.  Using NA QI solutions as initial points for boundary optimization will allow the expansion of the quasi-isodynamic library of suitable reactor configurations. The use of optimization within the NA space is being currently explored to identify particularly good initial points for further optimization. 

\section*{Acknowledgments}

The authors would like to thank Matt Landreman for providing the numerical code, described in \citet{plunk2019direct}, which was adapted for the present study. We are grateful to 
Carolin Nührenberg for providing the QIPC configuration and for illuminating clarifications, to Michael Drevlak for providing the code used for calculating the effective ripple, to Eduardo Rodríguez for pointing out the analytical conditions necessary for the helicity transition and for fruitful discussions, and to Per Helander for guidance while prepairing the manuscript.   This work was partly supported by a grant from the Simons Foundation (560651, KCM).

\section{Appendix I: Conditions for zeros of curvature}\label{Appendix I}

Given the need to have points of zero curvature on the magnetic axis, it is important to be able to control this behaviour at different orders. We reproduce here the relevant parts of Appendix B in \citep{camacho2022direct} for our discussion. 

We use the Fourier axis representation \ref{eq:R-Fourier} and \ref{eq:z-Fourier}, and we employ a Taylor expansion to describe the axis locally, around points of stellarator symmetry (which coincide with extrema of $B_0$),
\begin{eqnarray}
    R(\phi) = \sum_{i=0} \frac{R_{2i}}{(2i)!} \phi^{2i},\label{R-series}\\
    z(\phi) = \sum_{i=0} \frac{z_{2i+1}}{(2i+1)!} \phi^{2i+1},\label{z-series}
\end{eqnarray}

where $R_n$ and $Z_n$ denote the $n$-th derivatives with respect to $\varphi$. We can also see from the curvature and torsion expressions for a general parametrization, 
\begin{eqnarray}
    \kappa = \frac{|{\bf x}^\prime \times {\bf x}^{\prime\prime}|}{|{\bf x}^\prime|^3},\label{eq:curv-general}\\
    \tau = \frac{({\bf x}^\prime\times{\bf x}^{\prime\prime})\cdot{\bf x}^{\prime\prime\prime}}{|{\bf x}^\prime\times{\bf x}^{\prime\prime}|^2},
\end{eqnarray}
that we need to calculate up to third derivatives of the axis position function $x(\phi)$. These derivatives can be easily calculated by noting $d\hat{\bf R}/d\phi = \hat{\bf \phi}$ and $d\hat{\bf \phi}/d\phi = -\hat{\bf R}$. Further substituting the expansions for $R$ and $z$, Equations~\ref{R-series}-\ref{z-series}, the contributions to each derivative can be collected by their order in $\phi$, the following  properties can be confirmed
\begin{eqnarray}
    \hat{\bf R}\cdot\left(\frac{d^n{\bf x}}{d\phi^n}\right)_m = 0\text{, for odd } n + m,\\
    \hat{\bf z}\cdot\left(\frac{d^n{\bf x}}{d\phi^n}\right)_m = \hat{\bf \phi}\cdot\left(\frac{d^n{\bf x}}{d\phi^n}\right)_m = 0\text{, for even } n + m.
\end{eqnarray}
where the subscript $m$ denotes the (m+1)-order in $\phi$.

Now we can proceed to find the necessary conditions for having zeros in curvature at a given order (m+1). We will classify the zeros in curvature by the order of the first non-zero term in the power series, for example 
\begin{equation}
\kappa = \kappa_m \phi^m + \kappa_{m+1} \phi^{m+1} +\dots.
\end{equation}

Assuming that ${\bf x}^\prime$ is itself non-zero, the conditions on zeros in curvature are found by requiring \begin{equation}
{\bf x}^\prime \times {\bf x}^{\prime\prime} = 0,
\end{equation}
 at each order in $\phi$.  At first order ($m=0$), ${\bf x}^{\prime\prime}$ has its only non-zero component in the $\hat{\bf R}$ direction, and the condition ${\bf x}^\prime \times {\bf x}^{\prime\prime} = 0$ is satisfied by ${\bf x}^{\prime\prime}\cdot\hat{\bf R} = 0$, which results in the condition
 \begin{equation}
  R_0 - R_2 = 0.   
 \end{equation}

 The curvature can be made zero to higher order by considering higher-order contributions to ${\bf x}^{\prime\prime}$.  At odd orders, these are contained in the $\hat{\bf{ \phi}}-\hat{\bf z}$ plane and must be made parallel to the zeroth order contribution from ${\bf x}^\prime$, while at even orders, the even order contribution to ${\bf x}^{\prime\prime}$ must simply be zero. Thus, conditions at arbitrary order can be obtained, and a few are listed in Table \ref{curvature-table}.

\begin{table}
\begin{center}
\begin{tabular}{c c}
\hline
 Order & Constraint \\
\hline
 $0$ & $R_2=R_0$ \\
 $1$ & $z_3=2 z_1$ \\
 $2$ & $R_4=5 R_0$ \\
 $3$ & $z_5=16 z_1$ \\
 $4$ & $R_6=61 R_0$ \\
\hline
\end{tabular}
\caption{Conditions for zero curvature.}
\label{curvature-table}
\end{center}
\end{table}

The tabulated constraints on the derivatives of the axis components can be applied to a truncated Fourier representation.  Equations \ref{eq:Axis_Truncated_FourierRepresentation} are simply substituted into the constraint equations with $\phi$ set to a location of stellarator symmetry (for instance at $\phi = 0, \pi/N$ in the first period).  This results in a set of linear conditions on the Fourier coefficients that can be solved numerically, or by computer algebra.

As a simple example let us consider the family of 2-mode curves, 

\begin{eqnarray}
     R = 1 + R_c(1) \cos(N \phi) + R_c(2) \cos(2 N \phi),\\
     z = z_s(1) \sin(N \phi) + z_s(2) \sin(2 N \phi).
\end{eqnarray}
Enforcing the conditions for first order zeros of curvature ($R_0-R_2 = 0$) at extrema of $B_0$, $\phi = 0$ and $\phi = \pi/N$, gives
\begin{eqnarray}
R_c(1) = 0,\\
R_c(2) = -\frac{1}{4 N^2+1}.
\end{eqnarray}
Having second order zeros of $\kappa$ requires imposing conditions on the $z$ coefficients. At $\phi=0$ this results in 
\begin{eqnarray}
z_s(2) =  -\frac{2+N^2}{4+8N^2}z_s(1),
\end{eqnarray}
and when imposing it at the minima of $B_0$
\begin{eqnarray}
z_s(2) =  \frac{2+N^2}{4+8N^2}z_s(1),
\end{eqnarray}
 

\bibliographystyle{jpp}
\bibliography{omnigenity-near-axis}

\end{document}